\documentclass[prd,twocolumn,floatfix,amsmath,nofootinbib,amssymb,floatfix]{revtex4}
\usepackage{graphicx,color,dcolumn,booktabs,bm,multirow}
\usepackage{longtable,lscape}
\usepackage{txfonts}
\usepackage{overpic}
\usepackage{amssymb}
\usepackage{indentfirst}
\usepackage{feynmf}   %{feynmp}
\usepackage{slashed}  %for Feynman symbols
\usepackage{cases}
\usepackage{color}
\usepackage{multirow}
\usepackage{epstopdf}
\usepackage{csquotes}
\usepackage{graphicx,color,dcolumn,booktabs,bm}

\usepackage[colorlinks, citecolor=blue,anchorcolor=red,menucolor=red, linkcolor=red,filecolor=red,runcolor=red,urlcolor=blue,frenchlinks=red]{hyperref}

\begin{document}
%\begin{CJK}{GBK}{}
\title{Strong decays of the $DK^\ast$ and $\bar{D}K^{\ast}$ molecular states}
\author{Zi-Li Yue$^{1,2}$}
\author{Cheng-Jian Xiao$^3$}
\author{H. Garc{\'i}a-Tecocoatzi$^4$}
\author{Dian-Yong Chen$^{1,5}$\footnote{Corresponding author}} \email{chendy@seu.edu.cn}
\author{Elena Santopinto$^2$}\email{elena.santopinto@ge.infn.it}
\affiliation{
 $^{1}$ School of Physics, Southeast University,  Nanjing 210094, China}
\affiliation{
 $^{2}$ INFN, Sezione di Genova, Via Dodecaneso 33, 16146 Genova, Italy }
\affiliation{
 $^{3}$ Institute of Applied Physics and Computational Mathematics, Beijing 100088, China} 
\affiliation{
$^{4}$ Tecnologico de Monterrey, Escuela de Ingenieria y Ciencias, General Ramon Corona 2514,
Zapopan 45138, Mexico}
\affiliation{$^5$Lanzhou Center for Theoretical Physics, Lanzhou University, Lanzhou 730000, P. R. China}
\begin{abstract}
Inspired by the abundant structure near the threshold of the $D^{(*)}K^{(*)}/\bar{D}^{(*)}K^{(*)}$, we estimate the strong decay properties of the $T_{c\bar{s}1}^{f/a}$ and $T_{\bar{c}\bar{s}1}^{f/a}$ with $I(J^{P})=0/1(1^{+})$ in $DK^{*}$ and $\bar{D}K^{*}$ molecular scenarios in the present paper. By employing the effective Lagrangian approach, the widths of the processes $T_{c\bar{s}1}^{f}\to D^{*}K, D_{s}^{*}\eta, DK\pi$, $T_{c\bar{s}1}^{a}\to D^{*}K, D_{s}^{*}\pi, DK\pi$, and $T_{\bar{c}\bar{s}1}^{f/a}\to\bar{D}^{*}K, \bar{D}K\pi$ are estimated. Considering the present estimations, we propose to search for $T_{c\bar{s}1}^{f/a}$ states in $D^{*}K$ and $D_{s}^{*}\pi/D_{s}^{*}\eta$ mass invariant spectra. Their ratios may serve as an important test of the molecular scenario.

\end{abstract}
%\pacs{}

\maketitle
%\end{CJK}

\section{Introduction}\label{sec:1}

In the heavy-flavor meson sector, a number of exotic states that contain only one heavy quark have been observed in the last two decades. $D_{s0}^{*}(2317)$ is the first candidate. In 2003 the BABAR Collaboration reported the observation of  $D_{s0}^{*}(2317)$ in the invariant mass spectrum of $D_{s}^{+}\pi^{0}$ in the process of $e^{+}e^{-}$ annihilation~\cite{BaBar:2003oey}. Later, the CLEO Collaboration confirmed the observation of $D_{s0}^{*}(2317)$ and observed a new structure, $D_{s1}(2460)$, with a mass close to $2.46~\mathrm{GeV}$ in the $D_{s}^{*+}\pi^{0}$ channel, with a statistical significance of $5.7~\sigma$~\cite{CLEO:2003ggt}. The isospin, spin and parity of $D_{s0}^{*}(2317)$ and $D_{s1}(2460)$ are consistent with the values $I(J^{P})=0(0^{+})$ and $0(1^{+})$, respectively~~\cite{ParticleDataGroup:2024cfk,BaBar:2003oey,CLEO:2003ggt}, and their widths are both less than $3.8~\mathrm{MeV}$~\cite{ParticleDataGroup:2024cfk}. These two states could be candidates for the $P$-wave charm-strange mesons. However, the observed masses of $D_{s0}^{*}(2317)$ and $D_{s1}(2460)$ are approximately $160~\mathrm{MeV}$ and $100~\mathrm{MeV}$, respectively, below the predicted masses of the corresponding $P$-wave charm-strange mesons in Refs.~\cite{Godfrey:1985xj,Godfrey:2003kg}. As the masses of these states are near the thresholds of $DK$ and $D^{*}K$, the molecular interpretations have also been proposed and discussed in the literatures~\cite{Guo:2006fu,Faessler:2007gv,Faessler:2007us,Cleven:2014oka,Xiao:2016hoa,Wu:2019vsy,Zhu:2019vnr,Liu:2022dmm,Yue:2023qgx,Liu:2023cwk,Kim:2023htt}. Alternative interpretations include compact tetraquark configurations~\cite{Terasaki:2003qa,Dmitrasinovic:2004cu,Maiani:2004vq,Chen:2016spr,Maiani:2024quj}, conventional P-wave charm-strange mesons~\cite{Colangelo:2003vg,Wei:2005ag,Wang:2006fg,Wang:2006zw,Matsuki:2011xp,Ke:2013zs} and P-wave $c\bar{s}$ core coupled to the $DK/D^{*}K$ channels~\cite{vanBeveren:2003kd,Simonov:2004ar}.

In 2020, the LHCb Collaboration reported evidence of two charm-strange resonances, $X_{0}(2900)$ (now called $T_{\bar{c}\bar{s}0}^{*}(2870)^{0}$~\cite{ParticleDataGroup:2024cfk}) with spin-0 and $X_{1}(2900)$ (now called $T_{\bar{c}\bar{s}1}^{*}(2900)^{0}$~\cite{ParticleDataGroup:2024cfk}) with spin-1, both composed of $\bar{c}du\bar{s}$, in the  $D^- K^+$ invariant mass distributions of the decay process $B^{+}\to D^{+}D^{-}K^{+}$~\cite{LHCb:2020bls,LHCb:2020pxc}. The masses and widths of the two states were reported as~\cite{LHCb:2020pxc}:
\begin{eqnarray}
X_{0}(2900):M&=&2.866\pm0.007\pm0.002~\mathrm{GeV/c^2},\nonumber\\
\Gamma&=&57\pm12\pm4~\mathrm{MeV},\nonumber\\
X_{1}(2900):M&=&2.904\pm0.005\pm0.001~\mathrm{GeV/c^2},\nonumber\\
\Gamma&=&110\pm11\pm4~\mathrm{MeV}.
\end{eqnarray}
Very recently, an analysis of $B^{-}\to D^{-}D^{0}K_{S}^{0}$ by the LHCb Collaboration confirmed the existence of $T_{\bar{c}\bar{s}0}^{*}(2870)^{0}$, in the $D^{0}K_{S}^{0}$ invariant mass distribution, with a significance of 5.3~$\sigma$, although there was no evidence found for $T_{\bar{c}\bar{s}1}^{*}(2900)^{0}$~\cite{LHCb:2024xyx}. 

In 2022, the LHCb Collaboration reported the first observation of a spin-0 doubly charged tetraquark candidate, $T_{c\bar{s}0}^{a}(2900)^{++}$~($c\bar{s}u\bar{d}$), and of  its neutral isospin partner, $T_{c\bar{s}0}^{a}(2900)^{0}$~($c\bar{s}\bar{u}d$), in the $D_{s}\pi$ mass invariant distribution of the $B$ meson decay processes, with significances $8.0~\sigma$ and $6.5~\sigma$, respectively~\cite{LHCb:2022sfr,LHCb:2022lzp}. The superscript $a$ in the notation indicates that the isospin of $T_{c\bar{s}0}^{a}(2900)^{0/++}$ is equal to 1. The masses and widths are~\cite{LHCb:2022lzp}:
\begin{eqnarray}
T_{c\bar{s}0}^{a}(2900)^{0}:M&=&2.892\pm0.014\pm0.015~\mathrm{GeV},\nonumber\\
\Gamma&=&0.119\pm0.026\pm0.013~\mathrm{GeV},\nonumber\\
T_{c\bar{s}0}^{a}(2900)^{++}:M&=&2.921\pm0.017\pm0.020~\mathrm{GeV},\nonumber\\
\Gamma&=&0.137\pm0.032\pm0.017~\mathrm{GeV},
\end{eqnarray}
which are in good agreement with each other. This isospin $1$ resonance is now named $T_{c\bar{s}0}^{*}(2900)$ by PDG~ \cite{ParticleDataGroup:2024cfk}. As $T_{\bar{c}\bar{s}0}^{*}(2870)^{0}$, $T_{\bar{c}\bar{s}1}^{*}(2900)^{0}$ and $T_{c\bar{s}0}^{*}(2900)$  have four different quark flavors and their masses are close to the  $\bar{D}^{*}K^{*}/D^{*}K^{*}$ thresholds, molecular interpretations~\cite{Liu:2020nil,Molina:2020hde,He:2020btl,Hu:2020mxp,Agaev:2020nrc,Mutuk:2020igv,Huang:2020ptc,Yue:2022mnf,Chen:2022svh,Agaev:2022eyk,Ke:2022ocs,Duan:2023lcj,Wang:2023hpp,Yue:2023qgx,Huang:2023fvj,Yu:2023avh,Yang:2024coj} were proposed. Moreover, they were also interpreted as tetraquark states in Refs.~\cite{Zhang:2020oze,Wang:2020xyc,Wang:2020prk,He:2020jna,Guo:2021mja,Yang:2024coj}.

Furthermore, in 2024, the LHCb Collaboration discovered a new exotic $T_{c\bar{s}}^{++}$ and its isospin partner $T_{c\bar{s}}^{0}$ with a mass of $(2327\pm13\pm13)~\mathrm{MeV}$ and a width of $(96\pm16^{+170}_{-23})~\mathrm{MeV}$ with a significance of more than $10~\sigma$, in the decay process $D_{s1}(2460)^{+}\to D_{s}^{+}\pi^{+}\pi^{-}$~\cite{LHCb:2024iuo}. The $T_{c\bar{s}0}(2327)$ spin-parity was found to be $J^{P}=0^{+}$ with a significance that exceeded $10~\sigma$. The mass of this state is close to the $DK$ threshold of $2363~\mathrm{MeV}$. In Ref.~\cite{Wang:2024fsz}, the authors employed a final-state interaction approach in order to investigate the amplitude of $D_{s1}(2460)^{+}\to D_{s}^{+}\pi^{+}\pi^{-}$, and their results supported the explanation of $T_{c\bar{s}0}(2327)$ as a $DK$ molecular state. The lattice calculation strongly supported the hadronic molecular picture, while disfavoring the compact tetraquark model~\cite{Gregory:2025ium}.

\begin{figure}[t]
\centering
\includegraphics[width=8.5cm]{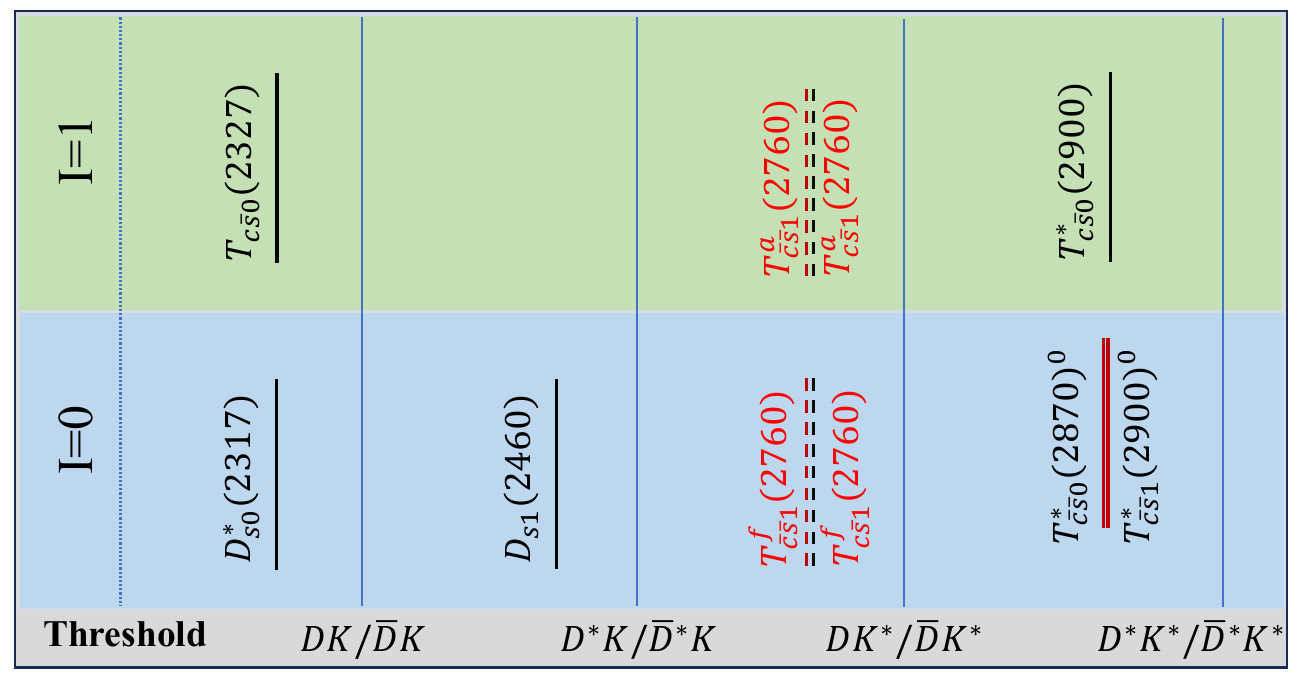}
\caption{The schematic picture of  states near the $D^{(\ast)}K^{(\ast)}/\bar{D}^{(\ast)}K^{(\ast)}$ thresholds. Here the solid black lines and solid red lines refer to the observed potential molecular states composed of $D^{(*)}K/D^{*}K^{*}$ and $\bar{D}^{*}K^{*}$, respectively, while the dashed lines indicate possible molecular states that have not yet been reported experimentally.} 
\label{Fig:Comp}
\end{figure}

As shown in Fig.~\ref{Fig:Comp}, the $DK$ molecular candidates $D_{s0}^\ast(2317)$ with $I=0$ and $T_{c\bar{s}0} (2327)$ with $I=1$ have been observed near the threshold of $DK$~\cite{BaBar:2003oey,CLEO:2003ggt,LHCb:2024iuo}. The $D^\ast K$ molecular candidate $D_{s1}(2460)$ has been observed near the $D^\ast K$ threshold~\cite{CLEO:2003ggt}, and the $D^{*}K^{*}/\bar{D}^{*}K^{*}$ molecular candidates $T_{\bar{c}\bar{s}0}^{*}(2870)^{0}$, $T_{\bar{c}\bar{s}1}^{*}(2900)^{0}$ and $T_{c\bar{s}0}^{*}(2900)$ have also been observed near the $D^{*}K^{*}/\bar{D}^{*}K^{*}$ thresholds~\cite{LHCb:2020bls,LHCb:2020pxc,LHCb:2022sfr,LHCb:2022lzp,LHCb:2024xyx}. However, the possible $DK^{*}/\bar{D}K^{*}$ molecular candidates predicted by numerous mass spectra theoretical studies~\cite{Kong:2021ohg, Yang:2023wgu, Hu:2020mxp, Wang:2023hpp, Chen:2023syh}, have not yet been observed experimentally. In Fig.~\ref{Fig:Comp} and in this article, we will refer to these possible $S-$wave $D K^\ast$ molecular states as $T_{c\bar{s}1}^{f}(2760)$ and  $T_{c\bar{s}1}^{a}(2760)$ with $I=0$ and $I=1$ respectively, and to the possible $S-$wave $\bar{D} K^\ast$ molecular states as $T_{\bar{c}\bar{s}1}^{f}(2760)$ and $T_{\bar{c}\bar{s}1}^{a}(2760)$ with $I=0$ and $I=1$ respectively. We use this $f/a$ notation for isospin $0/1$ in order to distinguish them from the PDG reported states~\cite{ParticleDataGroup:2024cfk}, while the mass in the notation is only used to indicate states near the $DK^{*}/\bar{D}K^{*}$ thresholds. In the present work, we investigate the the strong decay behaviors of $T_{c\bar{s}1}^{a/f}(2760)$ and $T_{\bar{c}\bar{s}1}^{a/f}(2760)$ in the molecular picture, by using an effective Lagrangian approach. The coupling between the molecular structure and their components was determined by the Weinberg's compositeness condition, which was originally proposed to study the deuteron as bound state of proton  and neutron~\cite{Weinberg:1962hj}. Then this method is extensively used to investigate of deuteron-like molecular states~\cite{Dong:2017gaw,Guo:2017jvc,Faessler:2007gv,Faessler:2007us}.

This paper is organized as follows. The hadronic molecular structures of $T_{c\bar{s}1}^{f/a}(2760)$ and $T_{\bar{c}\bar{s}1}^{f/a}(2760)$ states are introduced in Section \ref{sec:2}. The decay widths of the involved processes are estimated in Section \ref{sec:3}. The numerical results and the relevant discussions are presented in Section \ref{sec:4}, and the last section provides a short summary.

\section{Hadronic molecular structure}
\label{sec:2}

In this paper, we interpret $T_{c\bar{s}1}^{f/a}(2760)$ and $T_{\bar{c}\bar{s}1}^{f/a}(2760)$ (abbreviated to $T_{c\bar{s}1}^{f/a}$ and $T_{\bar{c}\bar{s}1}^{f/a}$ hereafter) as S-wave $DK^{*}$ and $\bar{D}K^{*}$ molecular states with quantum numbers $I(J^{P})=0/1(1^+)$. For the $D/\bar{D}$ and $K^\ast$ molecular states, the coordinates of $D/\bar{D}$ and $K^\ast$ can be expressed by the center of mass coordinate $x$ and their relative coordinate $y$, which are $x-\omega_{K^\ast D}y$ and $x+\omega_{DK^\ast} y$, respectively, with $\omega_{K^\ast D}=m_{K^\ast}/(m_{K^\ast}+m_D)$ and $\omega_{D K^\ast}=m_{D}/(m_{K^\ast}+m_D)$. Then, the effective Lagrangians describing the interaction between the $S-$wave molecular states and their constituents are:
\begin{eqnarray}
\label{eq:1}
\mathcal{L}_{T_{c\bar{s}1}}(x)&=&g_{T_{c\bar{s}1}}T_{c\bar{s}1}^{\mu+}(x)\int dy\Phi_{T_{c\bar{s}1}}(y^2)\left(D^{+}(x-\omega_{K^*D}y)\right.\nonumber\\&\times&\left.K^{*0}_{\mu}(x+\omega_{DK^*}y)\pm D^{0}(x-\omega_{K^*D}y)K^{*+}_{\mu}(x+\omega_{DK^*}y)\right)\nonumber\\&+&\text{H.c.},\nonumber\\
\mathcal{L}_{T_{\bar{c}\bar{s}1}}(x)&=&g_{T_{\bar{c}\bar{s}1}}T_{\bar{c}\bar{s}1}^{\mu+}(x)\int dy\Phi_{T_{\bar{c}\bar{s}1}}(y^2)\left(D^{-}(x-\omega_{K^*\bar{D}}y)\right.\nonumber\\&\times&\left.K^{*+}_{\mu}(x+\omega_{\bar{D}K^*}y)\pm \bar{D}^{0}(x-\omega_{K^*\bar{D}}y)K^{*0}_{\mu}(x+\omega_{\bar{D}K^*}y)\right)\nonumber\\&+&\text{H.c.},\label{Eq:Lag0}
\end{eqnarray}
where $\pm$ refer to the $T_{c\bar{s}1}$ states with $I=0$ ($T_{c\bar{s}1}^{f}$) and $I=1$ ($T_{c\bar{s}1}^{a}$), respectively, while the notation for the $T_{\bar{c}\bar{s}1}$ is the opposite. It worth mentioning that in the above effective Lagrangian, a correlation function $\Phi(y^2)$ depending on the relative coordinate $y$ is introduced to depicts the distributions of the $D/\bar{D}$ and $K^\ast$ componets in the molecular states. The Fourier transformation of the $\Phi(y^2)$ is:
\begin{equation}
\Phi(y^2)=\int \frac{d^4q}{(2\pi)^4}e^{-ipy}\tilde\Phi\left(-p^2\right).
\label{eq:fourier}
\end{equation}
A correlation function $\tilde\Phi\left(-p^2\right)$ should be chosen to describe the inner structure of the molecular states and should drop fairly rapidly in the ultraviolet region. Here, we adopt the correlation function in the Gaussian form~\cite{Faessler:2007gv, Faessler:2007us,Xiao:2020ltm,Chen:2016byt,Dong:2017gaw,Gutsche:2010jf}, which is:
\begin{equation}
\tilde\Phi(p_E^2,\Lambda^2)=\mathrm{exp}\left(-p_E^2/\Lambda^2\right), \label{Eq:CF}
\end{equation}
where $p_E$ is the Euclidean Jacobi momentum and $\Lambda$ is a model parameter that represents the distribution of the constituent mesons in the molecular states.

We can determine the coupling constants $g_{T_{c\bar{s}1}}/g_{T_{\bar{c}\bar{s}1}}$ in the effective Lagrangians in Eq.~\eqref{eq:1} by employing the compositeness condition~\cite{Weinberg:1962hj, Salam:1962ap, Hayashi:1967,vanKolck:2022lqz}:
\begin{eqnarray}
\label{eq:2}
Z&=&1-\Pi^{\prime}(m^2),\label{Eq:CP}
\end{eqnarray}
where the renormalization constant of the hadron wave function is equal to zero. $\Pi^{\prime}(m^2)$ is the derivative of the transverse part of the mass operator $\Pi^{\mu\nu}$:
\begin{equation}
\Pi^{\mu\nu}(p)=g^{\mu\nu}_{\perp}
\Pi(p^2)+\frac{p^{\mu}p^{\nu}}{p^2}
\Pi^{L}(p^2),
\end{equation}
with $\Pi(p^2)$ and $\Pi^L(p^2)$ being the conventional transverse and the longitudinal parts of the mass operator, respectively, and $g^{\mu\nu}_{\perp}=g^{\mu\nu}-p^{\mu}p^{\nu}/p^2$, $g_{\perp}^{\mu\nu}p_{\mu}=0$.

\begin{figure}[t]
\begin{tabular}{cc}
  \centering
  \includegraphics[width=4.2cm]{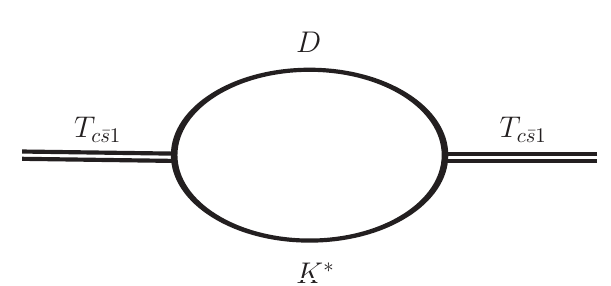}&
  \includegraphics[width=4.2cm]{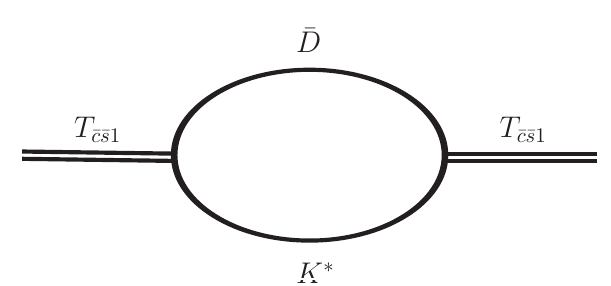}\\
\\
 $(a)$ & $(b)$ \\
\end{tabular}
\caption{The mass operators of $T_{c\bar{s}1}^{f/a}$ and $T_{\bar{c}\bar{s}1}^{f/a}$.\label{fig:MO}}
\end{figure}

Based on the effective Lagrangians in Eq.~\eqref{eq:1}, the concrete forms of the mass operators of $T_{c\bar{s}1}$ and $T_{\bar{c}\bar{s}1}$ corresponding to Fig.~\ref{fig:MO} are:
\begin{eqnarray}
\Pi^{\mu\nu}(m_{T_{c\bar{s}1}}^{2})&=&g_{T_{c\bar{s}1}}^{2} \Pi_0^{\mu\nu}(m_{T_{c\bar{s}1}}^{2}) \nonumber\\
\Pi^{\mu\nu}(m_{T_{\bar{c}\bar{s}1}}^{2})&=&g_{T_{\bar{c}\bar{s}1}}^{2} \Pi_0^{\mu\nu}(m_{T_{\bar{c}\bar{s}1}}^{2})
\end{eqnarray}
with 
\begin{eqnarray}
\Pi^{\mu\nu}_0(m_{T_{c\bar{s}1}}^{2})
&=&\int\frac{d^{4}q}{(2\pi)^{4}}\tilde{\Phi}^{2}\left[-(q-\omega_{DK^{*}}P)^{2},\Lambda^{2}\right]\nonumber\\
&\times&\frac{-g^{\mu\nu}+\left(p^{\mu}-q^{\mu}\right)\left(p^{\nu}-q^{\nu}\right)/m_{K^{*2}}}{(p-q)^{2}-m_{K^{*2}}}\frac{1}{q^{2}-m_{D}^{2}},\nonumber\\
\Pi_0^{\mu\nu}(m_{T_{\bar{c}\bar{s}1}}^{2})&=&\int\frac{d^{4}q}{(2\pi)^{4}}\tilde{\Phi}^{2}\left[-(q-\omega_{\bar{D}K^{*}}P)^{2},\Lambda^{2}\right]\nonumber\\
&\times&\frac{-g^{\mu\nu}+(p^{\mu}-q^{\mu})(p^{\nu}-q^{\nu})/m_{K^{*2}}}{(p-q)^{2}-m_{K^{*2}}}\frac{1}{q^{2}-m_{\bar{D}}^{2}}.\nonumber\\
\end{eqnarray}
As the masses of $D$ and $\bar{D}$ are the same, the values of the mass operators $\Pi^{\mu\nu}(m_{T_{c\bar{s}1}}^{2})$ and $\Pi^{\mu\nu}(m_{T_{\bar{c}\bar{s}1}}^{2})$ are the same, which lead to the same coupling constants $g_{T_{c\bar{s}1}}$ and $g_{T_{\bar{c}\bar{s}1}}$.

\begin{figure*}[htb]
\begin{tabular}{cccc}
  \centering
\includegraphics[width=4.2cm]{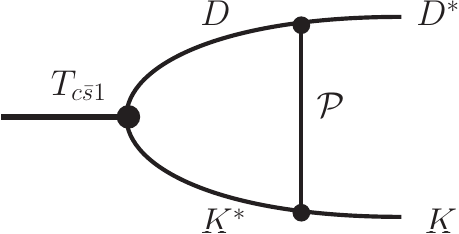}&
\includegraphics[width=4.2cm]{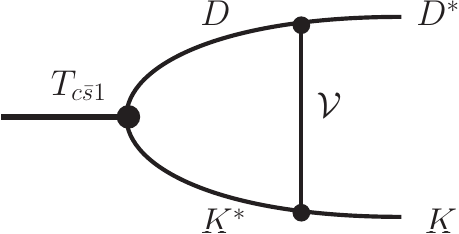}&
\includegraphics[width=4.2cm]{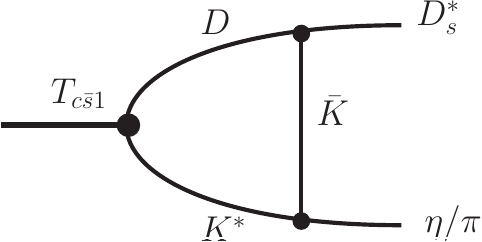}&
\includegraphics[width=4.2cm]{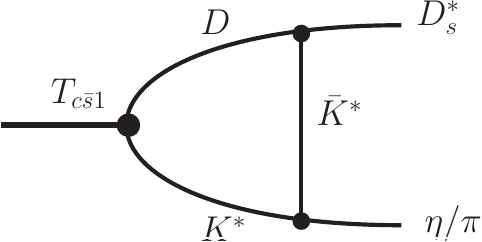}\\ \\
 $(a)$ & $(b)$ & $(c)$ & $(d)$ \\ \\
\includegraphics[width=4.2cm]{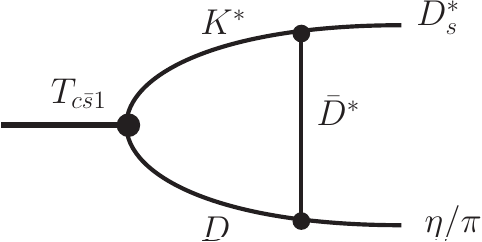}&
\includegraphics[width=4.2cm]{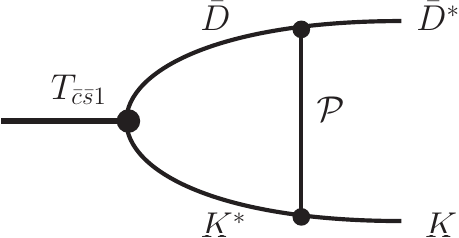}&
\includegraphics[width=4.2cm]{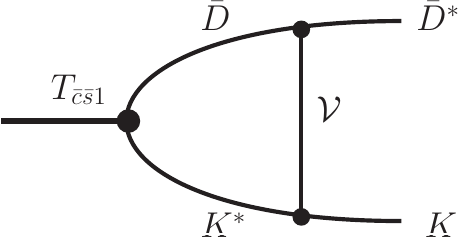}&
\includegraphics[width=4.2cm]{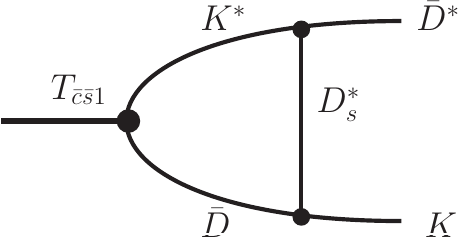}\\ \\
 $(e)$ & $(f)$ & $(g)$ & $(h)$ 
\end{tabular}
\caption{The typical diagrams contributing to $T_{c\bar{s}1}\to D^{*}K$ (diagrams (a)-(b)), $T_{c\bar{s}1}\to D_{s}^{*}\pi/\eta$ (diagrams (c)-(e)) and $T_{\bar{c}\bar{s}1}\to\bar{D}^{*}K$ (diagrams (f)-(h)).}
\label{fig:2}
\end{figure*}

\section{Strong decays of $T_{c\bar{s}1}^{f/a}$ and $T_{\bar{c}\bar{s}1}^{f/a}$}
\label{sec:3}

\renewcommand\arraystretch{1.4}
\begin{table}
\caption{The strong decay modes of $T_{c\bar{s}1}^{f/a}$ and $T_{\bar{c}\bar{s}1}^{f/a}$ involved in the present estimations. \label{Tab:DM}}
\begin{tabular}{p{2cm}<\centering p{6cm}<\centering }
\toprule[1pt]
States & Decay modes \\	
\midrule[1pt]
$T_{c\bar{s}1}^{f}$ & $ D^{*}K$,\  $D_{s}^{*}\eta$,\  $DK\pi$\\ 
$T_{c\bar{s}1}^{a}$ & $ D^{*}K$, \ $D_{s}^{*}\pi$, \ $DK\pi$\\
\midrule[1pt]
$T_{\bar{c}\bar{s}1}^{f}$ & $ \bar{D}^{*}K$,  \ $\bar{D}K\pi$\\ 
$T_{\bar{c}\bar{s}1}^{a}$ & $ \bar{D}^{*}K$,  \ $\bar{D}K\pi$\\
\bottomrule[1pt]
\end{tabular}	
\end{table}

In the present paper, $T_{c\bar{s}1}^{f/a}$ and $T_{\bar{c}\bar{s}1}^{f/a}$ are considered to be molecular states composed of $DK^{*}$ and $\bar{D}K^{*}$ with $I=0$ and $1$, respectively. The possible strong decay modes of $T_{c\bar{s}1}^{f/a}$ and $T_{\bar{c}\bar{s}1}^{f/a}$ are shown in Table~\ref{Tab:DM}. In the present estimations, these decay processes are estimated at the hadron level and the corresponding diagrams are depicted in Fig.~\ref{fig:2}.

Based on the heavy-quark limit and chiral symmetry, the effective Lagrangians describing interactions between charmed (charmed-strange) meson pairs and vector/pseudoscalar light mesons are~\cite{Kaymakcalan:1983qq, Oh:2000qr, Casalbuoni:1996pg, Colangelo:2002mj, He:2019csk, Ding:2008gr, Wu:2021udi, Liu:2020ruo}:
\begin{eqnarray}
\label{eq:3}
%\small
\mathcal{L}_{\mathcal{D}^{(*)}\mathcal{D}^{(*)}\mathcal{V}}&=&-ig_{\mathcal{D}\mathcal{D}V}\mathcal{D}_i^{\dagger}
   \overleftrightarrow{\partial}_{\mu}\mathcal{D}^j(\mathcal{V}^{\mu})^i_j-2f_{\mathcal{D}^*\mathcal{D}V}\epsilon_{\mu\nu\alpha\beta}\nonumber\\
   && \times(\partial^{\mu}\mathcal{V}^{\nu})^i_j
     (\mathcal{D}^{\dagger}_i\overleftrightarrow{\partial}^{\alpha}\mathcal{D}^{*\beta j}
     -\mathcal{D}_i^{*\beta\dagger}\overleftrightarrow{\partial}^{\alpha}\mathcal{D}^j)\nonumber\\
   &&+ig_{\mathcal{D}^*\mathcal{D}^*V}\mathcal{D}_i^{*\nu\dagger}\overleftrightarrow{\partial}_{\mu}
     \mathcal{D}_{\nu}^{*j}(\mathcal{V}^{\mu})^i_j\nonumber\\
    &&+4if_{\mathcal{D}^*\mathcal{D}^*V}\mathcal{D}_{i\mu}^{*\dagger}(\partial^{\mu}\mathcal{V}^{\nu}-\partial^{\nu}
     \mathcal{V}^{\mu})^i_j\mathcal{D}_{\nu}^{*j},\nonumber\\
\mathcal{L}_{\mathcal{D}^{(*)}\mathcal{D}^{(*)}\mathcal{P}}&=&-ig_{\mathcal{D}^*\mathcal{D}P}(\bar{\mathcal{D}}\partial_{\mu}\mathcal{P}\mathcal{D}^{*\mu}-\bar{\mathcal{D}}^{*\mu}\partial_{\mu}{\mathcal P}\mathcal{D})\nonumber\\
&&+\frac{1}{2}g_{\mathcal{D}^*\mathcal{D}^*P}\epsilon_{\mu\nu\alpha\beta}\bar{\mathcal{D}}^{*\mu}\partial_{\nu}\mathcal{P}
\overleftrightarrow{\partial}^{\alpha}\mathcal{D}^{*\beta},
\end{eqnarray}
where $\mathcal{D}^{(*)}=(D^{(*)0},D^{(*)+},D^{(*)+}_s)$ and $A\overleftrightarrow{\partial}B=A\partial B-B\partial A$. {$\mathcal{P}$} and {$\mathcal{V}$} are the matrix forms of the vector nonet and pseudo-scalar nonet, and their concrete forms are:
\begin{eqnarray}
\mathcal{P} &=&
\left(
\begin{array}{ccc}
\frac{\pi^{0}}{\sqrt 2}+\alpha\eta+\beta\eta\prime &\pi^{+} &K^{+}\\
\pi^{-} &-\frac{\pi^{0}}{\sqrt2}+\alpha\eta+\beta\eta \prime&K^{0}\\
K^{-} &\bar K^{0} &\gamma\eta+\delta\eta\prime
\end{array}
\right), \nonumber\\ \nonumber\\
\nonumber \\
%\end{eqnarray}
%
%\begin{eqnarray}
\mathcal{V} &=& \left(
\begin{matrix}
\frac{1}{\sqrt2}(\rho^0+\omega)&\rho^{+}&K^{*+}\\
\rho^{-}&\frac{1}{\sqrt2}(-\rho^{0}+\omega)&K^{*0}\\
K^{*-}&\bar K^{*0}&\phi
\end{matrix}
\right).
\end{eqnarray}

Where the corresponding mixing angles are defined as:\\

\begin{eqnarray}
\alpha &=&\frac{\mathrm{cos}\theta-\sqrt{2}\mathrm{sin}\theta}{\sqrt{2}}, \  \ \ \ \ \beta=\frac{\mathrm{sin}\theta+\sqrt{2}\mathrm{cos}\theta}{\sqrt{6}},\nonumber\\ \gamma &=&\frac{-2\mathrm{cos}\theta-\sqrt{2}\mathrm{sin}\theta}{\sqrt{6}}, \ \ \ \ \delta=\frac{-2\mathrm{sin}\theta+\sqrt{2}\mathrm{cos}\theta}{\sqrt{6}	},
\end{eqnarray}
here, we adopt $\theta=-19.1^{\circ}$~\cite{MARK-III:1988crp,Morgan:1970yz}.

According to the SU$(3)$ symmetry, the effective Lagrangians describing the interaction between light pseudoscalar mesons and vector mesons are~\cite{Liu:2005jb, Bando:1985rf,Oh:2000qr,Chen:2011cj,Haglin:2000ar,Lin:1999ad,Kaymakcalan:1983qq}:

\begin{eqnarray}
\mathcal{L}_{K^{*}KP}&=&-ig_{K^{*}KP}\Big(\bar{K}\partial^{\mu}P-\partial^{\mu}\bar{K}P\Big)+\mathrm{h.c.}\nonumber\\
\mathcal{L}_{K^{*}K^{*}P}&=&-g_{K^{*}K^{*}P}\epsilon_{\mu\nu\alpha\beta}\partial^{\alpha}\bar{K}^{*\beta}\partial^{\mu}K^{*\nu}P\nonumber\\
\mathcal{L}_{K^{*}KV}&=&-g_{K^{*}KV}\epsilon_{\mu\nu\alpha\beta}\partial^{\mu}\bar{K}^{*\nu}\partial^{\alpha}V^{\beta}K+\mathrm{h.c.},
\end{eqnarray}
where $P$ stands for the triplet of $\pi$, $\eta$ and $\eta^{\prime}$ from pseudoscalar nonet, and $V$ stands for the $\rho$ triplet and $\omega$ from the vector nonet. $\bar{K}^{*}$ is a doublet of $\bar{K}^{(*)}=\Big(K^{(*)-},\bar{K}^{(*)0}\Big)$. The coupling constants involved in the above effective Lagrangians will be discussed in the following section.

%\begin{widetext}
\subsection{Two-body decay process}

Using the effective Lagrangians listed above, we can obtain the amplitudes corresponding to the Feynman diagrams in Fig.~\ref{fig:2}-(a)-(e), which are:
\begin{eqnarray}
i\mathcal{M}_{a}&=&i^{3}\int\frac{d^{4}q}{(2\pi)^{4}}\Big[g_{T_{c\bar{s}1}}\tilde{\Phi}(-p_{12}^{2},\Lambda^{2})\epsilon_{\mu}(p)\Big]\nonumber\\
&\times& \Big[-ig_{D^{*}Dp}(iq^{\theta})\epsilon_{\theta}(p_{3})\Big]\Big[ig_{K^{*}Kp}(-iq^{\nu}-ip^{\nu}_{4})\Big]\nonumber\\&\times&\frac{1}{p_{1}^{2}-m_{1}^{2}}\frac{1}{q^{2}-m_{q}^{2}}\frac{-g^{\mu\nu}+p_{2}^{\mu}p_{2}^{\nu}/m_{2}^{2}}{p_{2}^{2}-m_{2}^{2}},\nonumber\\
%\end{eqnarray*}
%\begin{eqnarray*}
i\mathcal{M}_{b}&=&i^{3}\int\frac{d^{4}q}{(2\pi)^{4}}\Big[g_{T_{c\bar{s}1}}\tilde{\Phi}(-p_{12}^{2},\Lambda^{2})\epsilon^{\tau}(p)\Big]\nonumber\\&\times&\Big[-2f_{D^{*}DV}\epsilon_{\mu\nu\alpha\beta}iq^{\mu}(ip_{3}^{\alpha}+ip_{1}^{\alpha})\epsilon^{\beta}(p_{3})\Big]\nonumber\\&\times&\Big[g_{K^{*}KV}\epsilon_{\omega\delta\rho\sigma}(-iq^{\omega})(-ip_{2}^{\rho})\Big]\frac{1}{p_{1}^{2}-m_{1}^{2}}\nonumber\\&\times&\frac{-g_{\tau\sigma}+p_{2\tau}p_{2\sigma}/m_{2}^{2}}{p_{2}^{2}-m_{2}^{2}}\frac{-g_{\nu\delta}+q_{\nu}q_{\delta}/m_{q}^{2}}{q^{2}-m_{q}^{2}},\nonumber\\
%\end{eqnarray*}
%\begin{eqnarray}
i\mathcal{M}_{c}&=&i^{3}\int\frac{d^{4}q}{(2\pi)^{4}}\Big[g_{T_{c\bar{s}1}}\tilde{\Phi}(-p_{12}^{2},\Lambda^{2})\epsilon_{\mu}(p)\Big]\nonumber\\&\times&\Big[-ig_{D^{*}DP}(iq^{\theta})\epsilon_{\theta}(p_{3})\Big]\Big[ig_{K^{*}KP}(ip_{4\nu}+iq_{\nu})\Big]\nonumber\\&\times&\frac{1}{p_{1}^{2}-m_{1}^{2}}\frac{1}{q^{2}-m_{q}^{2}}\frac{-g^{\mu\nu}+p_{2}^{\mu}p_{2}^{\nu}/m_{2}^{2}}{p_{2}^{2}-m_{2}^{2}},\nonumber\\
i\mathcal{M}_{d}&=&i^{3}\int\frac{d^{4}q}{(2\pi)^{4}}\Big[g_{T_{c\bar{s}1}}\tilde{\Phi}(-p_{12}^{2},\Lambda^{2})\epsilon^{\tau}(p)\Big]\nonumber\\&\times&\Big[-2f_{D^{*}DV}\epsilon_{\mu\nu\alpha\beta}(iq^{\mu})(ip_{3}^{\alpha}+ip_{1}^{\alpha})\epsilon^{\beta}(p_{3})\Big]\nonumber\\&\times&\Big[g_{K^{*}K^{*}P}\epsilon_{\omega\delta\rho\sigma}(-ip_{2}^{\omega})(-iq^{\rho})\Big]\frac{1}{p_{1}^{2}-m_{1}^{2}}\nonumber\\&\times&\frac{-g_{\tau\delta}+p_{2\tau}p_{2\delta}/m_{2}^{2}}{p_{2}^{2}-m_{2}^{2}}\frac{-g_{\nu\sigma}+q_{\nu}q_{\sigma}/m_{q}^{2}}{q^{2}-m_{q}^{2}},\nonumber
\end{eqnarray}
\begin{eqnarray}
i\mathcal{M}_{e}&=&i^{3}\int\frac{d^{4}q}{(2\pi)^{4}}\left[g_{T_{c\bar{s}1}}\tilde{\Phi}(-p_{12}^{2},\Lambda^{2})\epsilon_{\phi}(p)\right]\nonumber\\&\times&\epsilon^{\theta}(p_{3})\Big[ig_{D^{*}D^{*}V}(ip_{3\delta}-iq_{\delta})g_{\lambda\theta}\Big.\nonumber\\&+&\Big.4if_{D^{*}D^{*}V}(-ip_{1\lambda}g_{\theta\delta}+ip_{1\theta}g_{\lambda\delta})\Big]\nonumber\\&\times&\Big[-ig_{D^{*}DP}(ip_{4\xi})\Big]\frac{-g^{\phi\delta}+p_{1}^{\phi}p_{1}^{\delta}/m_{1}^{2}}{p_{1}^{2}-m_{1}^{2}}\nonumber\\&\times&\frac{1}{p_{2}^{2}-m_{2}^{2}}\frac{-g^{\lambda\xi}+q^{\lambda}q^{\xi}/m_{q}^{2}}{q^{2}-m_{q}^{2}}.
\end{eqnarray}
%\end{widetext}
It should be noted that the correlation function introduced by the effective Lagrangian in Eq.~\eqref{Eq:Lag0} is involved in the above amplitudes. As indicated in Eq.~\eqref{Eq:CF}, the correlations function in the Gaussian form ensures all the loop integrals finite in the ultraviolet region.

The total contributions for $T_{c\bar{s}1}\to D^{*+}K^{0}$ and $T_{c\bar{s}1}\to D_{s}^{*+}\eta/\pi^{0}$ are:
\begin{eqnarray}
\label{eq:twoboday-1}
\mathcal{M}_{T_{c\bar{s}1}\to D^{*+}K^{0}}&=&\mathcal{M}_{a}^{\pi^{0}}+\mathcal{M}_{a}^{\eta}+\mathcal{M}_{a}^{\eta^{\prime}}+\mathcal{M}_{b}^{\rho^{0}}+\mathcal{M}^{\omega}_{b}\nonumber\\&\pm&\left(\mathcal{M}_{a}^{\pi^{-}}+\mathcal{M}_{b}^{\rho^{-}}\right),\nonumber\\
\mathcal{M}_{T_{c\bar{s}1}^{f}\to D_{s}^{*}\eta}&=&\mathcal{M}_{c}^{\bar{K}^{0}}+\mathcal{M}_{d}^{\bar{K}^{*0}}+\mathcal{M}_{e}^{D^{*-}}+\mathcal{M}_{c}^{K^{-}}+\mathcal{M}_{d}^{K^{*-}}\nonumber\\&+&\mathcal{M}_{e}^{\bar{D}^{*0}},\nonumber\\
\mathcal{M}_{T_{c\bar{s}1}^{a}\to D_{s}^{*}\pi^{0}}&=&\mathcal{M}_{c}^{\bar{K}^{0}}+\mathcal{M}_{d}^{\bar{K}^{*0}}+\mathcal{M}_{e}^{D^{*-}}-\left(\mathcal{M}_{c}^{K^{-}}+\mathcal{M}_{d}^{K^{*-}}\right.\nonumber\\&+&\left.\mathcal{M}_{e}^{\bar{D}^{*0}}\right),
\end{eqnarray}
where $\pm$ corresponds to initial states $T_{c\bar{s}1}^{f}$ and $T_{c\bar{s}1}^{a}$, respectively. 

The amplitudes for $T_{\bar{c}\bar{s}1}\to \bar{D}^{*}K$ presented in Fig.~\ref{fig:2} (diagrams (f)-(h)) can be easily obtained by substituting the meson masses and coupling constants in the amplitudes $\mathcal{M}_{a}$, $\mathcal{M}_{b}$ and $\mathcal{M}_{e}$. 

The total amplitudes for the processes $T_{\bar{c}\bar{s}1}^{f/a}\to D^{*-}K^{+}$ are:
%%%%%% 
\begin{eqnarray}
\label{eq:twoboday-2}
\mathcal{M}_{T_{\bar{c}\bar{s}1}\to D^{*-}K^{+}}&=&\mathcal{M}_{a}^{\pi^{0}}+\mathcal{M}_{a}^{\eta}+\mathcal{M}_{a}^{\eta^{\prime}}+\mathcal{M}_{b}^{\rho^{0}}+\mathcal{M}_{b}^{\omega}\nonumber\\&\pm&\left(\mathcal{M}_{a}^{\pi^{+}}+\mathcal{M}_{b}^{\rho^{+}}+\mathcal{M}_e^{D_{s}^{*}}\right),
\end{eqnarray}
where $\pm$ refers to initial states $T_{\bar{c}\bar{s}1}^a$ and $T_{\bar{c}\bar{s}1}^f$, respectively.

With the total amplitudes for the-two body decay processes of $T_{c\bar{s}1}$ and $T_{\bar{c}\bar{s}1}$ in Eq.~(\ref{eq:twoboday-1}) and (\ref{eq:twoboday-2}), we can calculate the partial widths of the considered processes by means of:
\begin{eqnarray}
\Gamma_{T_{c\bar{s}1}/T_{\bar{c}\bar{s}1}\to...}=\frac{1}{3}\frac{1}{8\pi}\frac{|\vec{p}|}{M^{2}}\overline{|\mathcal{{M}}_{T_{c\bar{s}1}/T_{\bar{c}\bar{s}1}\to...}|^{2}},
\end{eqnarray}
where the overline indicates the sum of the spin of the states involved, while the factor $1/3$ comes from the spin average of the initial state.

For the processes $T_{c\bar{s}1}\to D^{*}K$ and $T_{\bar{c}\bar{s}1} \to \bar{D}^\ast K$, the partial decay width of $D^{*+}K^{0}$ ($D^{*-}K^{+}$) and $D^{*0}K^{+}$ ($\bar{D}^{*0}K^{0}$) channels make the same contribution. Therefore:
\begin{eqnarray}
\Gamma(T_{c\bar{s}1}\to D^{*}K)&\approx&2\Gamma(T_{c\bar{s}1}\to D^{*+}K^{0}),\nonumber\\
\Gamma(T_{\bar{c}\bar{s}1}\to \bar{D}^{*}K)&\approx&2\Gamma(T_{\bar{c}\bar{s}1}\to {D}^{*-}K^{+}).
\end{eqnarray}

\subsection{Three-body decay process}
\begin{figure}[t]
  \centering
  \includegraphics[width=6cm]{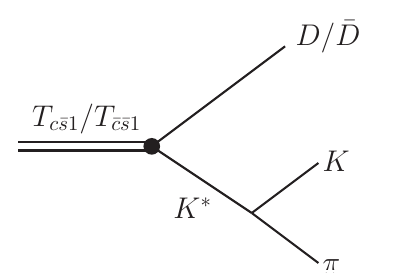}
  \caption{The three-body decay of $T_{c\bar{s}1}^{f/a}$ and $T_{\bar{c}\bar{s}1}^{f/a}$ .}\label{fig:3}
\end{figure}

In addition to two body decay processes, 
we also consider three body decays of $T_{c\bar{s}1}$ and $T_{\bar{c}\bar{s}1}$ since the width of $K^\ast$ is around 50 MeV, which dominantly decay into $K\pi$, the amplitude corresponding to the diagram  shown in Fig.~\ref{fig:3} reads:
\begin{eqnarray}
\mathcal{M}&=&\Big[g_{T_{c\bar{s}1}}\tilde{\Phi}(-p_{12}^{2},\Lambda^{2})\epsilon^{\mu}(p)\Big]\Big[ig_{K^{*}KP}i(p_{3}^{\nu}-p_{2}^{\nu})\Big]\nonumber\\&\times&\frac{-g_{\mu\nu}+q_{\mu}q_{\nu}/m_{K^{*}}^{2}}{(p-p_{1})^{2}-m_{K^{*}}^{2}+im_{K^{*}}\Gamma_{K^{*}}},
\end{eqnarray}
then the partial widths for the three body decay channels could be estimated by:
\begin{eqnarray}
d\Gamma=\frac{1}{3} \frac{1}{(2\pi)^{3}}\frac{1}{32M^{3}}\overline{|{\mathcal{M}}|^{2}}dm_{12}^{2}dm_{23}^{2}.
\end{eqnarray}

\section{Numerical Results and discussions}
\label{sec:4}
\subsection{Coupling Constants}

\begin{figure}[t]
  \centering
  \includegraphics[width=8.3 cm]{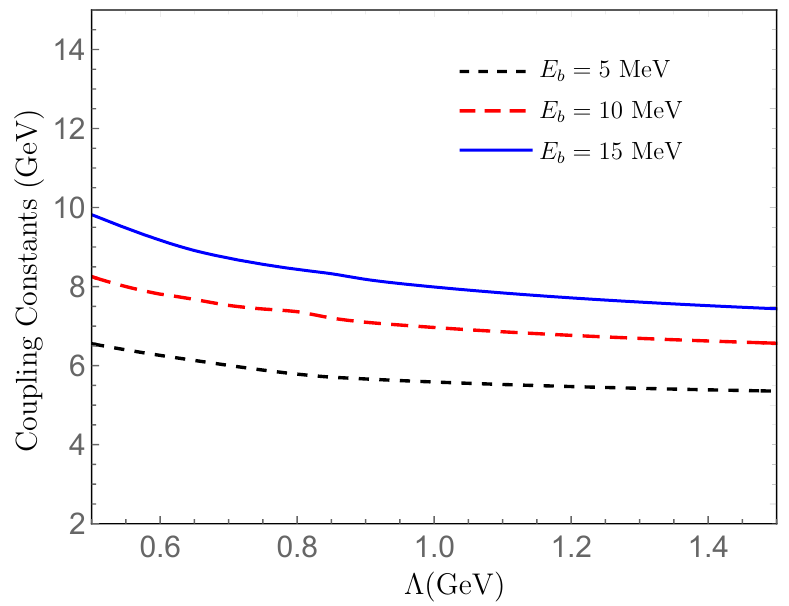}
\caption{(Color online.) The coupling constants $g_{T_{c\bar{s}1}} /g_{T_{\bar{c}\bar{s}1}}$ depending on model parameter $\Lambda$ with $E_{b}=5~\mathrm{MeV}$, $10~\mathrm{MeV}$ and $15~\mathrm{MeV}$.}\label{fig:4}
\end{figure}  

Before the estimations of the partial widths of the processes involved, the relevant coupling constants should be further clarified. The coupling constants related to the molecular states and their components could be estimated by employing Weinberg's compositeness in Eq.~(\ref{eq:2}). To date, the $T_{c\bar{s}1}^{f/a}/T_{\bar{c}\bar{s}1}^{f/a}$ have not been observed experimentally, here, we take three typical values as binding energys of these possible states: $5~\mathrm{MeV}$, $10~\mathrm{MeV}$ and $15~\mathrm{MeV}$. In addition, the model parameter $\Lambda$ introduced by the correlation function should be in the order of $1~\mathrm{GeV}$. In this paper, we vary $\Lambda$ from $0.5~\mathrm{GeV}$ to $1.5~\mathrm{GeV}$. The coupling constants $g_{T_{c\bar{s}1}}/g_{T_{\bar{c}\bar{s}1}}$ depending on model parameter $\Lambda$ and binding energy $E_{b}$ are presented in Fig.~\ref{fig:4}. The figure shows that a larger binding energy corresponds to a larger coupling constant with the same parameter $\Lambda$. Specifically, in the parameter range considered the couplings $g_{T_{c\bar{s}1}}/g_{T_{\bar{c}\bar{s}1}}$ decrease from $6.55$ to $5.35$, from $8.25$ to $6.57$ and from $9.82$ to $7.44$ for $E_{b}=5~\mathrm{MeV}$, $E_{b}=10~\mathrm{MeV}$ and $E_{b}=15~\mathrm{MeV}$, respectively. It is worth noting that in the small bounding limit, the coupling constant $g_{T_{c\bar{s}1}}/g_{T_{\bar{c}\bar{s}1}}$ could be estimated by~\cite{Baru:2003qq, Lin:2017mtz},
\begin{eqnarray}
	g^2_{T_{c\bar{s}1}}= g^2_{T_{\bar{c}\bar{s}1}} = 4\pi \frac{(m_D+m_{K^\ast})^{5/2}}{(m_D m_{K^\ast})^{1/2}} \sqrt{32 E_b}, 
\end{eqnarray}
then the coupling constant was estimated to be 6.70, 7.96, 8.81 for $E_b=5$, $10$, and $15$ MeV, respectively, which are consistent with the present estimations within the uncertainties.

The coupling constants between charm-meson pairs and light mesons could be related to the same couplings in the heavy-quark limit and chiral symmetry by:
\begin{eqnarray}
g_{\mathcal{D}^{*}\mathcal{D}\mathcal{P}}&=&\frac{2g}{f_{\pi}}\sqrt{m_{\mathcal{D}}m_{\mathcal{D}^{*}}},~~~~g_{\mathcal{D}^{*}\mathcal{D}^{*}\mathcal{P}}=\frac{2g}{f_{\pi}},\nonumber\\
g_{\mathcal{D}\mathcal{D}\mathcal{V}}&=&\frac{\beta_{0}g_{v}}{\sqrt{2}},~~~~ f_{\mathcal{D}^{*}\mathcal{D}\mathcal{V}}=\frac{\lambda g_{v}}{\sqrt{2}},\nonumber\\
g_{\mathcal{D}^{*}\mathcal{D}^{*}\mathcal{V}}&=&\frac{\beta_{0}g_{v}}{\sqrt{2}},~~~~f_{\mathcal{D}^{*}\mathcal{D}^{*}\mathcal{V}}=\frac{\lambda g_{v}}{\sqrt{2}}m_{\mathcal{D}^{*}},
\end{eqnarray}
 where $f_{\pi}=132~\mathrm{MeV}$ is the decay constant of the pion, $g_{v}=m_{\rho}/f_{\pi}$, $\beta_{0}=0.9$ and $\lambda=0.56~\mathrm{GeV}^{-1}$~\cite{Kaymakcalan:1983qq, Oh:2000qr, Casalbuoni:1996pg, Aubert:2006mh}. $g=0.59$ is estimated by the process $D^{*+}\to D^{+}\pi^{0}$ in experiment~\cite{ParticleDataGroup:2024cfk,Belyaev:1994zk}.

In SU$(3)$ symmetry, the couplings $g_{K^{*}KP}$, $g_{K^{*}K^{*}P}$ and $g_{K^{*}KV}$ related to the same gauge-coupling constant $g$ are:
\begin{eqnarray}
g_{K^{*}KP}&=&\frac{1}{4}g,\nonumber\\
g_{K^{*}K^{*}P}&=&g_{K^{*}KV}=\frac{1}{4}\frac{g^2N_{c}}{16\pi^2f_{\pi}},
\end{eqnarray}
where $Nc=3$ is the color degree of freedom. $g=12.8$ is determined by the measured width of $K^{*0}\to K^{0}\pi^{0}$~\cite{ParticleDataGroup:2024cfk}.

\subsection{$T_{c\bar{s}1}^{f/a}$ decay properties}

\begin{figure}[t]
\centering
\includegraphics[width=8.5cm]{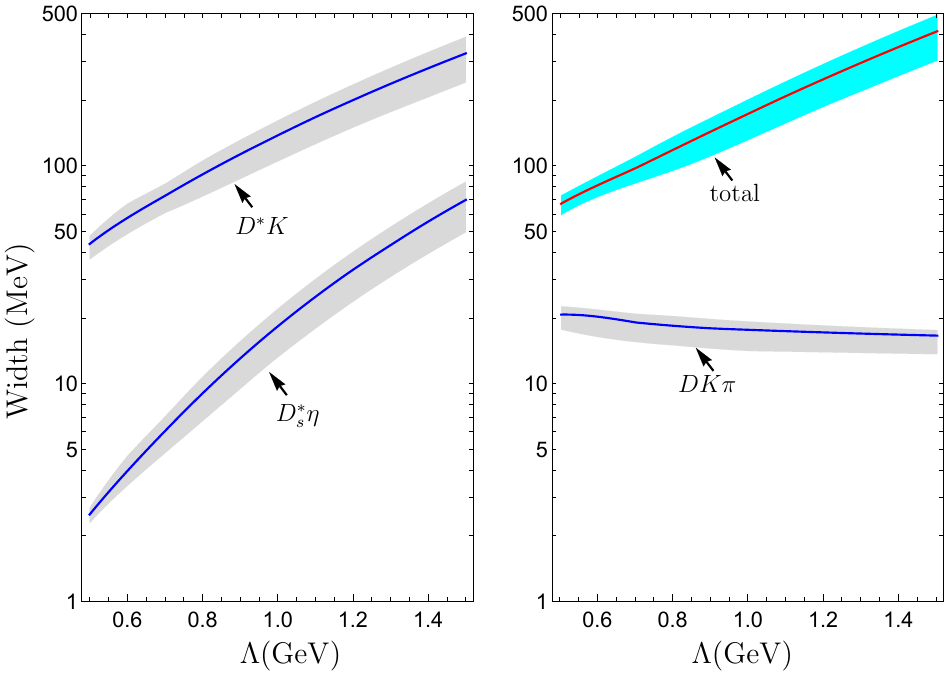}
\caption{(Color online.) The partial widths of $T_{c\bar{s}1}^{f}\to D^{*}K$, $D_{s}^{*}\eta$ and $DK\pi$  depending on the model parameter $\Lambda$. The total width is the sum of the widths of the three channels considered. The middle blue/red curve is the estimations corresponding to $E_{b}=10~\mathrm{MeV}$. The upper and lower limits of the gray/cyan band are the results of $E_b=15~\mathrm{MeV}$ and $E_b=5~\mathrm{MeV}$, respectively. \label{Fig:Tcsbarf}}
\end{figure}

\begin{figure}[t]
\centering
\includegraphics[width=8.5cm]{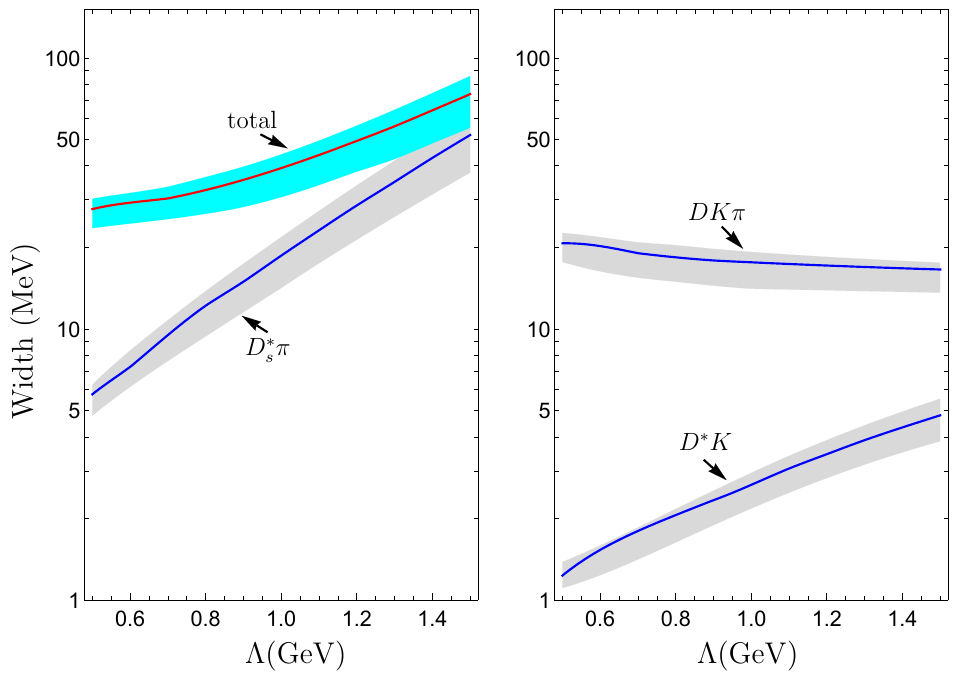}
\caption{(Color online.) The partial widths of $T_{c\bar{s}1}^{a}\to D^{*}K$, $D_{s}^{*}\pi$ and $DK\pi$ depending on the model parameter $\Lambda$ and binding energy $E_{b}$. The total width is the sum of the widths of the three channels considered. \label{Fig:Tcsbara}}
\end{figure}

In Fig.~\ref{Fig:Tcsbarf}, we present the estimations of the processes $T_{c\bar{s}1}^{f}\to D^{*}K$, $D_{s}^{*}\eta$, $DK\pi$ and the total width of these channels depending on parameter $\Lambda$ and binding energy $E_{b}$. The results indicate that $D^{*}K$ is the main decay channel of $T_{c\bar{s}1}^{f}$. Specifically, the partial decay width of the process $T_{c\bar{s}1}^{f}\to D^{*}K$ varies from $43.6~\mathrm{MeV}$ to $327.15~\mathrm{MeV}$ in the parameter range considered. The widths varies significantly with the parameter $\Lambda$. Therefore, we introduce the ratio of $\Gamma_{T_{c\bar{s}1}^f\to D_{s}^{*}\eta}$ to $\Gamma_{T_{c\bar{s}1}^f\to D^{*}K}$:

\begin{eqnarray}
    \frac{\Gamma_{T_{c\bar{s}1}^f\to D_s^\ast \eta}}{\Gamma_{T_{c\bar{s}1}^f\to D^\ast K}} = 0.06\sim 0.21.
\end{eqnarray}

Our estimations of the partial and total widths of $T_{c\bar{s}1}^a$ are presented in Fig.~\ref{Fig:Tcsbara}. As the case of $T_{c\bar{s}1}^f$, the partial widths into $D_s^\ast \pi$ and $D^\ast K$ strongly dependent on the model parameter $\Lambda$. However, regarding the $T_{c\bar{s}1}^a$, the width of $D_s^\ast \pi$ is much larger than that of $D^\ast K$. Specifically, in the parameter range considered, the partial widths of the $D_s^\ast \pi $ and $D^\ast K$ channels are estimated to be $(5.73\sim 51.8)$ and $(1.23\sim 4.80)$ MeV , respectively, and their ratio is estimated to be:
\begin{eqnarray}
    \frac{\Gamma_{T_{c\bar{s}1}^a\to D_s^\ast \pi}}{\Gamma_{T_{c\bar{s}1}^a\to D^\ast K}} =4.65\sim 10.8. 
\end{eqnarray}

Regarding $T_{c\bar{s}1}^f \to DK\pi$, the decay width is the same as that of $T_{c\bar{s}1}^a \to DK\pi$ in the $DK^\ast$ molecular frame, namely $20.7 \sim 16.5$ MeV in the parameter range considered.

\begin{figure}[t]
\centering
\includegraphics[width=8.5cm]{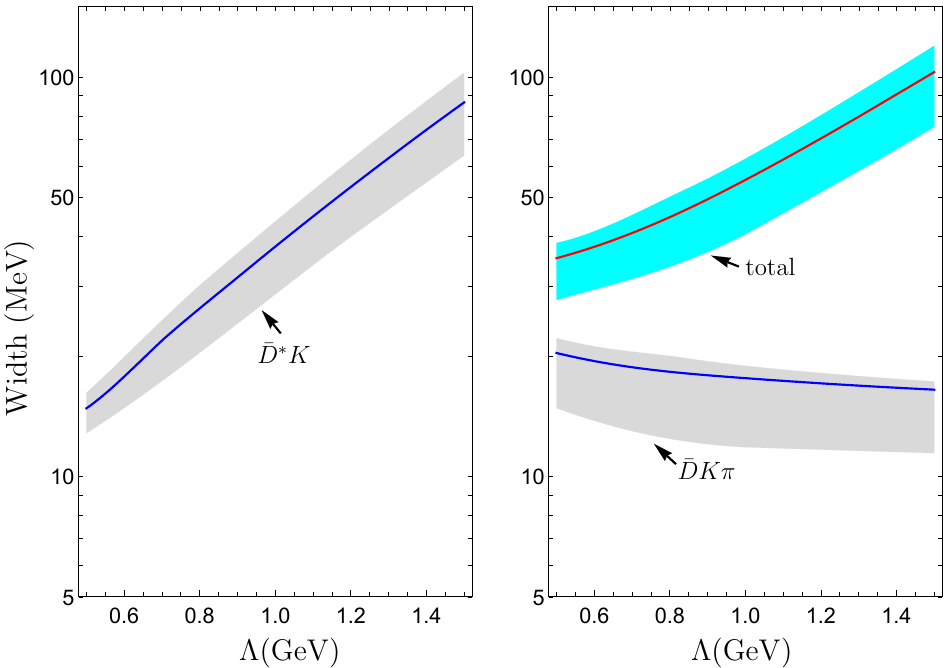}
\caption{(Color online.) The $\Lambda$ and $E_{b}$ dependence of the partial widths of $T_{\bar{c}\bar{s}1}^{f}\to\bar{D}^{*}K$, $\bar{D}K\pi$. The total width is the sum of the  widths of the two decay channels considered. \label{Fig:Tcbarsbarf}}
\end{figure}

\begin{figure}[t]
\centering
\includegraphics[width=8.5cm]{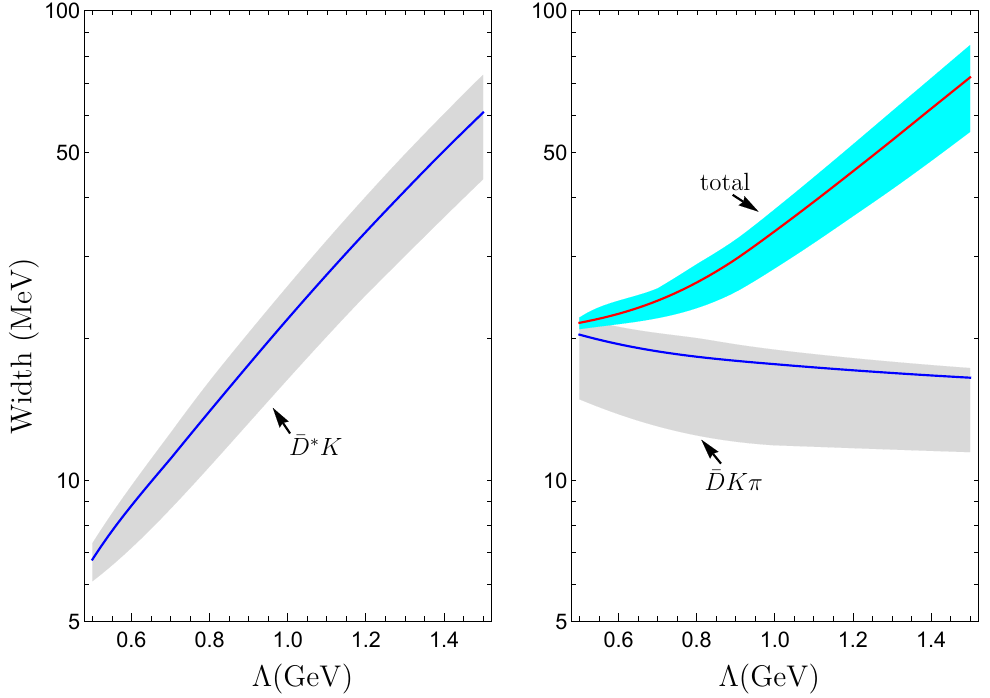}
\caption{(Color online.) The same as Fig.~\ref{Fig:Tcbarsbarf} but for $T_{\bar{c}\bar{s}1}^{a}$. \label{Fig:Tcbarsbara}}
\end{figure}

\begin{table}[htb]
   \centering
\caption{The total widths of $T_{c\bar{s}1}^{f/a} /T_{\bar{c}\bar{s}1}^{f/a}$, and the branching fractions of the processes considered, estimated in the $DK^\ast/\bar{D}K^{*} $ molecular frame with $E_b=10$ MeV.}
    \label{tab:my_label}    
   \begin{tabular}{cccc}
\toprule[1pt]    
States & Total Width (MeV) & Decay Modes & Branching ratio ($\%$)\\
\midrule[1pt]
 & & $D^\ast K$ & $[65.4-79.0]$\\
$T_{c\bar{s}}^f$ & $[66.8-413]$, & $D_s^\ast \eta$ &  $[3.80-16.8]$\\
 & & $D K \pi $ & $[30.8-4.20]$\\
 & & $D^\ast K$ & $[4.50-6.50]$\\
$T_{c\bar{s}}^a$ & $[27.7-73.1]$, & $D_s^\ast \pi$ &  $[20.8-70.9]$\\
 & & $D K \pi $ & $[74.7-22.6]$\\
\midrule[1pt]
\multirow{2}{*}{$T_{\bar{c}\bar{s}}^f$} & \multirow{2}{*}{$[35.1-103]$} & $\bar{D}^\ast K$ & $[47.3-80.9]$
\\
& & $\bar{D} K\pi $& $[52.7-19.1]$ \\
\multirow{2}{*}{$T_{\bar{c}\bar{s}}^a$} & \multirow{2}{*}{$[27.1-77.1]$} & $\bar{D}^\ast K$ & $[29.0-74.9]$\\
& & $\bar{D} K\pi $& $[71.0-25.1]$ \\
\bottomrule[1pt]
    \end{tabular}
\end{table}

\subsection{$T_{\bar{c}\bar{s}1}^{f/a}$ decay properties}

The partial and total decay widths for $T_{\bar{c}\bar{s}1}^{f}\to\bar{D}^{*}K, \bar{D}K\pi$ and $T_{\bar{c}\bar{s}1}^{a}\to\bar{D}^{*}K, \bar{D}K\pi$ depending on model parameter $\Lambda$ and the binding energy $E_{b}$ are presented in Fig.~\ref{Fig:Tcbarsbarf} and Fig.~\ref{Fig:Tcbarsbara}, respectively. Similar to the case of $T_{c\bar{s}1}^f$ and $T_{c\bar{s}1}^a$, the partial widths of the three-body decay processes $T_{\bar{c}\bar{s}1}^{f/a} \to \bar{D} K \pi$ are the same. Specifically, the partial width is estimated to be $20.3\sim 16.5$ MeV in the parameter range considered. As for the $\bar{D}^\ast K$ channel, the partial decay widths are estimated to be $14.8\sim 86.4$ MeV and $6.76\sim 60.6$ MeV for $T_{\bar{c}\bar{s} 1}^f$ and $T_{\bar{c}\bar{s} 1}^a$, respectively.

Table~\ref{tab:my_label} shows the total widths of $T_{c\bar{s} 1}^{f/a}/ T_{\bar{c}\bar{s} 1}^{f/a}$ and the branching fractions of the processes considered, as estimated in the $DK^\ast /\bar{D}K^{*}$ molecular frame with $E_b=10$ MeV. For the $T_{c\bar{s}1}^{f/a}$ states, the state with $I=0$, i.e., $T_{c\bar{s}1}^{f}$, is more likely to be observed in $D^{*}K$ channel, while $T_{c\bar{s}1}^{a}$ is more likely to be found in the $D_{s}^{*}\pi$ channel. Regarding the $T_{\bar{c}\bar{s}1}^{f/a}$ states, there is no significant difference between $T_{\bar{c}\bar{s}1}^{f}$ and $T_{\bar{c}\bar{s}1}^{a}$. Indeed, both $T_{\bar{c}\bar{s}1}^{f/a}$ states have a considerable branching ratio in both $\bar{D}^{*}K$ and $\bar{D}K\pi$, and so they can be observed in those two channels.

It is worth noting that in Ref.~\cite{LHCb:2024vfz} the mass invariant distribution of $D^{*-}K^{+}$ of the process $B^+ \to D^+ D^{\ast-} K^+$ is not well described around $2.76~\mathrm{GeV}$ (see Fig.1-(e) of  Ref.~\cite{LHCb:2024vfz}). The experimental analysis in Ref.~\cite{LHCb:2024vfz} indicates that \enquote{when including an additional resonance with the mass and width fixed to $2750$ MeV and 100 MeV, respectively, the largest log-likelihood value improvement is around 30 units when the quantum numbers are $J^P=1^+$}. This may serve as an experimental hint for the existence of $\bar{D}K^\ast$ molecular states, although further experimental data are required. On the theoretical side, the results of the present article indicate that the total widths of both $T_{\bar{c}\bar{s}1}^{f/a}$ could reach about 100 MeV (as from Table~\ref{tab:my_label} $\Gamma(T_{\bar{c}\bar{s}1}^{f})=35.1-103~\mathrm{MeV}$ and $\Gamma(T_{\bar{c}\bar{s}1}^{a})=27.1-77.1~\mathrm{MeV}$). Thus, the possible isospin quantum number of the structure of the $D^{*-}K^{+}$ mass invariant distribution in the  process $B^+ \to D^+ D^{\ast-} K^+$ cannot be inferred according to its width. Moreover, it could also be that both structures are present (Note that the $\bar{D}^{*}K$ channel in Table~\ref{tab:my_label} includes not only $D^{*-}K^{+}$, but also $\bar{D}^{*0}K^{0}$).

\section{summary}
\label{sec:5}
From the long-standing $D_{s0}^\ast(2317)$ and $D_{s1}(2460)$ to the newly observed $T_{\bar{c}\bar{s}0}^{*}(2870)^{0}$ and $T_{\bar{c}\bar{s}1}^{*}(2900)^{0}$, $T_{c\bar{s}0}^{*}(2900)$ and $T_{c\bar{s}0}(2327)$, the experimental observations indicate that the exotic hadron states near the thresholds of $DK$, $D^{\ast} K$ and $D^{\ast} K^\ast /\bar{D}^\ast K^\ast$ are abundant. The near-threshold feature of these states has prompted the widespread adoption of molecular state descriptions. Inspired by the experimental observations~\cite{LHCb:2024vfz} and the theoretical predictions~\cite{Kong:2021ohg, Yang:2023wgu, Chen:2023syh, Hu:2020mxp, Wang:2023hpp} of the existence of $D K^\ast$ and $\bar{D}K^\ast$ molecular states, in the present paper, we investigate the strong decay behaviors of the possible $S-$wave $DK^{*}$ and $\bar{D}K^{*}$ molecular states $T_{c\bar{s}1}^{f/a}$ and $T_{\bar{c}\bar{s}1}^{f/a}$ with quantum numbers $I(J^{P})=0/1(1^{+})$.

Through an effective Lagrangian approach, the widths of the possible decay channels, including $T_{c\bar{s}1}^{f}\to D^{*}K, D_{s}^{*}\eta, DK\pi$, $T_{c\bar{s}1}^{a}\to D^{*}K, D_{s}^{*}\pi, DK\pi$ and $T_{\bar{c}\bar{s}1}^{f/a}\to\bar{D}^{*}K, \bar{D}K\pi$, are estimated. Our estimations indicate that the ratio of $\Gamma_{T_{c\bar{s}1}^{f}\to D_{s}^{*}\eta}/\Gamma_{T_{c\bar{s}1}^{f}\to D^{*}K}$ is very different from $\Gamma_{T_{c\bar{s}1}^{a}\to D_{s}^{*}\pi}/\Gamma_{T_{c\bar{s}1}^{a}\to D^{*}K}$ in the $T_{c\bar{s}1}$ molecular scenario; this could serve as an important test of the molecular scenario. In addition, our calculations indicate that the widths of both $T_{\bar{c}\bar{s}1}^{f}$ and $T_{\bar{c}\bar{s}1}^{a}$ could be around 100 MeV, which is consistent with the experimental hint for the possible structure of the $D^{\ast-} K^+$ invariant mass distribution in the process $B^+ \to D^+ D^{\ast -} K^+$. More statistics and more data samples may shed light on the structure near the threshold of $DK^\ast/\bar{D}K^\ast$.

\section{ACKNOWLEDGMENTS} 
This study is partly supported by the National Natural Science Foundation of China under Grant Nos. 12175037 and 12335001, and is supported, in part, by National Key Research and Development Program under contract No. 2024YFA1610503. Zi-Li Yue is also supported by the SEU Innovation Capability Enhancement Plan for Doctoral Students (Grant No. CXJH$\_$SEU 24135) and the China Scholarship Council (Grant No. 202406090305).


\begin{thebibliography}{99}
\bibitem{BaBar:2003oey}
B.~Aubert \textit{et al.} [BaBar],
%``Observation of a narrow meson decaying to $D_s^+ \pi^0$ at a mass of 2.32-GeV/c$^2$,''
Phys. Rev. Lett. \textbf{90} (2003), 242001
doi:10.1103/PhysRevLett.90.242001
[arXiv:hep-ex/0304021 [hep-ex]].
%997 citations counted in INSPIRE as of 26 Mar 2024

\bibitem{CLEO:2003ggt}
D.~Besson \textit{et al.} [CLEO],
%``Observation of a narrow resonance of mass 2.46-GeV/c**2 decaying to D*+(s) pi0 and confirmation of the D*(sJ)(2317) state,''
Phys. Rev. D \textbf{68} (2003), 032002
[erratum: Phys. Rev. D \textbf{75} (2007), 119908]
doi:10.1103/PhysRevD.68.032002
[arXiv:hep-ex/0305100 [hep-ex]].
%708 citations counted in INSPIRE as of 26 Mar 2024

\bibitem{ParticleDataGroup:2024cfk}
S.~Navas \textit{et al.} [Particle Data Group],
%``Review of particle physics,''
Phys. Rev. D \textbf{110} (2024) no.3, 030001
doi:10.1103/PhysRevD.110.030001
%685 citations counted in INSPIRE as of 01 Jan 2025

\bibitem{Godfrey:1985xj}
S.~Godfrey and N.~Isgur,
%``Mesons in a Relativized Quark Model with Chromodynamics,''
Phys. Rev. D \textbf{32} (1985), 189-231
doi:10.1103/PhysRevD.32.189
%3205 citations counted in INSPIRE as of 26 Mar 2024

\bibitem{Godfrey:2003kg}
S.~Godfrey,
%``Testing the nature of the D(sJ)*(2317)+ and D(sJ)(2463)+ states using radiative transitions,''
Phys. Lett. B \textbf{568} (2003), 254-260
doi:10.1016/j.physletb.2003.06.049
[arXiv:hep-ph/0305122 [hep-ph]].
%186 citations counted in INSPIRE as of 26 Mar 2024

\bibitem{Guo:2006fu}
F.~K.~Guo, P.~N.~Shen, H.~C.~Chiang, R.~G.~Ping and B.~S.~Zou,
%``Dynamically generated 0+ heavy mesons in a heavy chiral unitary approach,''
Phys. Lett. B \textbf{641} (2006), 278-285
doi:10.1016/j.physletb.2006.08.064
[arXiv:hep-ph/0603072 [hep-ph]].
%278 citations counted in INSPIRE as of 26 Mar 2024

\bibitem{Faessler:2007gv}
A.~Faessler, T.~Gutsche, V.~E.~Lyubovitskij and Y.~L.~Ma,
%``Strong and radiative decays of the D(s0)*(2317) meson in the DK-molecule picture,''
Phys. Rev. D \textbf{76} (2007), 014005
doi:10.1103/PhysRevD.76.014005
[arXiv:0705.0254 [hep-ph]].
%207 citations counted in INSPIRE as of 26 Mar 2024

\bibitem{Faessler:2007us}
A.~Faessler, T.~Gutsche, V.~E.~Lyubovitskij and Y.~L.~Ma,
%``D* K molecular structure of the D(s1)(2460) meson,''
Phys. Rev. D \textbf{76} (2007), 114008
doi:10.1103/PhysRevD.76.114008
[arXiv:0709.3946 [hep-ph]].
%114 citations counted in INSPIRE as of 26 Mar 2024

\bibitem{Cleven:2014oka}
M.~Cleven, H.~W.~Grie\ss{}hammer, F.~K.~Guo, C.~Hanhart and U.~G.~Mei\ss{}ner,
%``Strong and radiative decays of the $D^*_{s0}(2317)$ and $D_{s1}(2460)$,''
Eur. Phys. J. A \textbf{50} (2014), 149
doi:10.1140/epja/i2014-14149-y
[arXiv:1405.2242 [hep-ph]].
%68 citations counted in INSPIRE as of 01 Jan 2025
%\cite{Gershon:2022xnn}

\bibitem{Xiao:2016hoa}
c.~J.~Xiao, D.~Y.~Chen and Y.~L.~Ma,
%``Radiative and pionic transitions from the $D_{s1}(2460)$ to the $D_{s0}^\ast(2317)$,''
Phys. Rev. D \textbf{93} (2016) no.9, 094011
doi:10.1103/PhysRevD.93.094011
[arXiv:1601.06399 [hep-ph]].
%26 citations counted in INSPIRE as of 26 Mar 2024

\bibitem{Wu:2019vsy}
T.~W.~Wu, M.~Z.~Liu, L.~S.~Geng, E.~Hiyama and M.~P.~Valderrama,
%``$DK$, $DDK$, and $DDDK$ molecules\textendash{}understanding the nature of the $D_{s0}^*(2317)$,''
Phys. Rev. D \textbf{100} (2019) no.3, 034029
doi:10.1103/PhysRevD.100.034029
[arXiv:1906.11995 [hep-ph]].
%51 citations counted in INSPIRE as of 01 Jan 2025

\bibitem{Zhu:2019vnr}
H.~Zhu and Y.~Huang,
%``Production of the $D_{s0}(2317)$ and $D_{s1}(2460)$ by kaon-induced reactions on a proton target,''
Phys. Rev. D \textbf{100} (2019) no.5, 054031
doi:10.1103/PhysRevD.100.054031
[arXiv:1904.06641 [hep-ph]].
%1 citations counted in INSPIRE as of 01 Jan 2025

\bibitem{Liu:2022dmm}
M.~Z.~Liu, X.~Z.~Ling, L.~S.~Geng, En-Wang and J.~J.~Xie,
%``Production of Ds0*(2317) and Ds1(2460) in B decays as D(*)K and Ds(*)\ensuremath{\eta} molecules,''
Phys. Rev. D \textbf{106} (2022) no.11, 114011
doi:10.1103/PhysRevD.106.114011
[arXiv:2209.01103 [hep-ph]].
%21 citations counted in INSPIRE as of 01 Jan 2025

\bibitem{Yue:2023qgx}
Z.~L.~Yue, C.~J.~Xiao and D.~Y.~Chen,
%``Pionic and radiative transitions from $T_{c\bar{s}0}^+(2900)$ to $D_{s1}^+(2460)$ as a probe of the structure of $D_{s1}^+(2460)$,''
Eur. Phys. J. C \textbf{83} (2023) no.8, 769
doi:10.1140/epjc/s10052-023-11948-3
[arXiv:2308.15355 [hep-ph]].
%2 citations counted in INSPIRE as of 01 Jan 2025

\bibitem{Liu:2023cwk}
M.~Z.~Liu, X.~Z.~Ling and L.~S.~Geng,
%``Productions of Ds0*(2317) and Ds1(2460) in B(s) and \ensuremath{\Lambda}b(\ensuremath{\Xi}b) decays,''
Phys. Rev. D \textbf{109} (2024) no.5, 5
doi:10.1103/PhysRevD.109.056014
[arXiv:2312.01433 [hep-ph]].
%4 citations counted in INSPIRE as of 01 Jan 2025

\bibitem{Kim:2023htt}
H.~J.~Kim and H.~C.~Kim,
%``D*s0(2317) and B*s0 as Molecular States,''
PTEP \textbf{2024} (2024) no.7, 073D01
doi:10.1093/ptep/ptae095
[arXiv:2310.13370 [hep-ph]].
%3 citations counted in INSPIRE as of 01 Jan 2025
%\cite{Aubert:2006mh}

\bibitem{Terasaki:2003qa}
K.~Terasaki,
%``BABAR resonance as a new window of hadron physics,''
Phys. Rev. D \textbf{68} (2003), 011501
doi:10.1103/PhysRevD.68.011501
[arXiv:hep-ph/0305213 [hep-ph]].
%168 citations counted in INSPIRE as of 14 May 2025

\bibitem{Dmitrasinovic:2004cu}
V.~Dmitrasinovic,
%``D*+(S) (2317) and D*+(S) (2460): Tetraquarks bound by the t Hooft instanton-induced interaction?,''
Phys. Rev. D \textbf{70} (2004), 096011
doi:10.1103/PhysRevD.70.096011
%33 citations counted in INSPIRE as of 14 May 2025
%\cite{Maiani:2004vq}

\bibitem{Maiani:2004vq}
L.~Maiani, F.~Piccinini, A.~D.~Polosa and V.~Riquer,
%``Diquark-antidiquarks with hidden or open charm and the nature of X(3872),''
Phys. Rev. D \textbf{71} (2005), 014028
doi:10.1103/PhysRevD.71.014028
[arXiv:hep-ph/0412098 [hep-ph]].
%962 citations counted in INSPIRE as of 14 May 2025

\bibitem{Chen:2016spr}
H.~X.~Chen, W.~Chen, X.~Liu, Y.~R.~Liu and S.~L.~Zhu,
%``A review of the open charm and open bottom systems,''
Rept. Prog. Phys. \textbf{80} (2017) no.7, 076201
doi:10.1088/1361-6633/aa6420
[arXiv:1609.08928 [hep-ph]].
%401 citations counted in INSPIRE as of 14 May 2025

\bibitem{Maiani:2024quj}
L.~Maiani, A.~D.~Polosa and V.~Riquer,
%``Open charm tetraquarks in broken SU(3)F symmetry,''
Phys. Rev. D \textbf{110} (2024) no.3, 034014
doi:10.1103/PhysRevD.110.034014
[arXiv:2405.08545 [hep-ph]].
%7 citations counted in INSPIRE as of 14 May 2025

\bibitem{Colangelo:2003vg}
P.~Colangelo and F.~De Fazio,
%``Understanding D(sJ)(2317),''
Phys. Lett. B \textbf{570} (2003), 180-184
doi:10.1016/j.physletb.2003.08.003
[arXiv:hep-ph/0305140 [hep-ph]].
%195 citations counted in INSPIRE as of 14 May 2025
%\cite{Matsuki:2011xp}

\bibitem{Wei:2005ag}
W.~Wei, P.~Z.~Huang and S.~L.~Zhu,
%``Strong decays of D(sJ)(2317) and D(sJ)(2460),''
Phys. Rev. D \textbf{73} (2006), 034004
doi:10.1103/PhysRevD.73.034004
[arXiv:hep-ph/0510039 [hep-ph]].
%46 citations counted in INSPIRE as of 14 May 2025
%\cite{Colangelo:2003vg}

\bibitem{Wang:2006fg}
F.~L.~Wang, X.~L.~Chen, D.~H.~Lu, S.~L.~Zhu and W.~Z.~Deng,
%``Decays of D*(sj)(2317) and D(sj)(2460) Mesons in the Quark Model,''
HEPNP \textbf{30} (2006), 1041-1047
[arXiv:hep-ph/0604090 [hep-ph]].
%9 citations counted in INSPIRE as of 14 May 2025

\bibitem{Wang:2006zw}
Z.~G.~Wang,
%``Structure of the axial-vector meson D(s1)(2460),''
J. Phys. G \textbf{34} (2007), 753-765
doi:10.1088/0954-3899/34/4/011
[arXiv:hep-ph/0611271 [hep-ph]].
%29 citations counted in INSPIRE as of 14 May 2025

\bibitem{Matsuki:2011xp}
T.~Matsuki and K.~Seo,
%``Chiral Particle Decay of Heavy-Light Mesons in a Relativistic Potential Model,''
Phys. Rev. D \textbf{85} (2012), 014036
doi:10.1103/PhysRevD.85.014036
[arXiv:1111.0857 [hep-ph]].
%20 citations counted in INSPIRE as of 14 May 2025
%\cite{Wang:2006fg}

\bibitem{Ke:2013zs}
H.~W.~Ke, X.~Q.~Li and Y.~L.~Shi,
%``The radiative decays of $0^{++}$ and $1^{+-}$ heavy mesons,''
Phys. Rev. D \textbf{87} (2013) no.5, 054022
doi:10.1103/PhysRevD.87.054022
[arXiv:1301.4014 [hep-ph]].
%20 citations counted in INSPIRE as of 14 May 2025

\bibitem{vanBeveren:2003kd}
E.~van Beveren and G.~Rupp,
%``Observed $D_s(2317)$ and tentative $D(2100\text{--}2300)$ as the
charmed cousins of the light scalar nonet,''
Phys. Rev. Lett. \textbf{91} (2003), 012003
doi:10.1103/PhysRevLett.91.012003
[arXiv:hep-ph/0305035 [hep-ph]].
%400 citations counted in INSPIRE as of 28 May 2025

\bibitem{Simonov:2004ar}
Y.~A.~Simonov and J.~A.~Tjon,
%``The Coupled-channel analysis of the D and D(s) mesons,''
Phys. Rev. D \textbf{70} (2004), 114013
doi:10.1103/PhysRevD.70.114013
[arXiv:hep-ph/0409361 [hep-ph]].
%72 citations counted in INSPIRE as of 28 May 2025

\bibitem{LHCb:2020bls}
R.~Aaij \textit{et al.} [LHCb],
%``A model-independent study of resonant structure in $B^+\to D^+D^-K^+$ decays,''
Phys. Rev. Lett. \textbf{125} (2020), 242001
doi:10.1103/PhysRevLett.125.242001
[arXiv:2009.00025 [hep-ex]].
%179 citations counted in INSPIRE as of 26 Mar 2024

\bibitem{LHCb:2020pxc}
R.~Aaij \textit{et al.} [LHCb],
%``Amplitude analysis of the $B^+\to D^+D^-K^+$ decay,''
Phys. Rev. D \textbf{102} (2020), 112003
doi:10.1103/PhysRevD.102.112003
[arXiv:2009.00026 [hep-ex]].
%215 citations counted in INSPIRE as of 26 Mar 2024

\bibitem{LHCb:2024xyx}
R.~Aaij \textit{et al.} [LHCb],
%``Observation of the Open-Charm Tetraquark Candidate Tcs0*(2870)0 in the B-\textrightarrow{}D-D0KS0 Decay,''
Phys. Rev. Lett. \textbf{134} (2025) no.10, 101901
doi:10.1103/PhysRevLett.134.101901
[arXiv:2411.19781 [hep-ex]].
%3 citations counted in INSPIRE as of 21 May 2025

\bibitem{LHCb:2022sfr}
R.~Aaij \textit{et al.} [LHCb],
%``First Observation of a Doubly Charged Tetraquark and Its Neutral Partner,''
Phys. Rev. Lett. \textbf{131} (2023) no.4, 041902
doi:10.1103/PhysRevLett.131.041902
[arXiv:2212.02716 [hep-ex]].
%94 citations counted in INSPIRE as of 31 Dec 2024

\bibitem{LHCb:2022lzp}
R.~Aaij \textit{et al.} [LHCb],
%``Amplitude analysis of B0\textrightarrow{}D\textasciimacron{}0Ds+\ensuremath{\pi}- and B+\textrightarrow{}D-Ds+\ensuremath{\pi}+ decays,''
Phys. Rev. D \textbf{108} (2023) no.1, 012017
doi:10.1103/PhysRevD.108.012017
[arXiv:2212.02717 [hep-ex]].
%42 citations counted in INSPIRE as of 26 Mar 2024

\bibitem{Liu:2020nil}
M.~Z.~Liu, J.~J.~Xie and L.~S.~Geng,
%``$X_0(2866)$ as a $D^*\bar{K}^*$ molecular state,''
Phys. Rev. D \textbf{102} (2020) no.9, 091502
doi:10.1103/PhysRevD.102.091502
[arXiv:2008.07389 [hep-ph]].
%78 citations counted in INSPIRE as of 26 Mar 2024

\bibitem{Molina:2020hde}
R.~Molina and E.~Oset,
%``Molecular picture for the $X_0(2866)$ as a $D^* \bar{K}^*$ $J^P=0^+$ state and related $1^+,2^+$ states,''
Phys. Lett. B \textbf{811} (2020), 135870
[erratum: Phys. Lett. B \textbf{837} (2023), 137645]
doi:10.1016/j.physletb.2020.135870
[arXiv:2008.11171 [hep-ph]].
%60 citations counted in INSPIRE as of 26 Mar 2024

\bibitem{He:2020btl}
J.~He and D.~Y.~Chen,
%``Molecular picture for $X_0(2900)$ and $X_1(2900)$,''
Chin. Phys. C \textbf{45} (2021) no.6, 063102
doi:10.1088/1674-1137/abeda8
[arXiv:2008.07782 [hep-ph]].
%51 citations counted in INSPIRE as of 01 Jan 2025

\bibitem{Hu:2020mxp}
M.~W.~Hu, X.~Y.~Lao, P.~Ling and Q.~Wang,
%``$X_0$(2900) and its heavy quark spin partners in molecular picture,''
Chin. Phys. C \textbf{45} (2021) no.2, 021003
doi:10.1088/1674-1137/abcfaa
[arXiv:2008.06894 [hep-ph]].
%50 citations counted in INSPIRE as of 26 Mar 2024

\bibitem{Agaev:2020nrc}
S.~S.~Agaev, K.~Azizi and H.~Sundu,
%``New scalar resonance X 0(2900) as a molecule: mass and width,''
J. Phys. G \textbf{48} (2021) no.8, 085012
doi:10.1088/1361-6471/ac0b31
[arXiv:2008.13027 [hep-ph]].
%52 citations counted in INSPIRE as of 01 Jan 2025
 %\cite{Weinberg:1962hj}

\bibitem{Mutuk:2020igv}
H.~Mutuk,
%``Monte-Carlo based QCD sum rules analysis of $X_0$(2900) and $X_1$(2900),''
J. Phys. G \textbf{48} (2021) no.5, 055007
doi:10.1088/1361-6471/abeb7f
[arXiv:2009.02492 [hep-ph]].
%26 citations counted in INSPIRE as of 01 Jan 2025

\bibitem{Huang:2020ptc}
Y.~Huang, J.~X.~Lu, J.~J.~Xie and L.~S.~Geng,
%``Strong decays of ${\bar{D}}^{*}K^{*}$ molecules and the newly observed $X_{0,1}$ states,''
Eur. Phys. J. C \textbf{80} (2020) no.10, 973
doi:10.1140/epjc/s10052-020-08516-4
[arXiv:2008.07959 [hep-ph]].
%66 citations counted in INSPIRE as of 01 Jan 2025

\bibitem{Yue:2022mnf}
Z.~L.~Yue, C.~J.~Xiao and D.~Y.~Chen,
%``Decays of the fully open flavor state Tcs\textasciimacron{}00 in a D*K* molecule scenario,''
Phys. Rev. D \textbf{107} (2023) no.3, 034018
doi:10.1103/PhysRevD.107.034018
[arXiv:2212.03018 [hep-ph]].
%17 citations counted in INSPIRE as of 26 Mar 2024

\bibitem{Chen:2022svh}
R.~Chen and Q.~Huang,
%``From the isovector molecular explanation of the newly $T_{c\bar{s}}^{a0(++)}(2900)$ to possible charmed-strange molecular pentaquarks,''
[arXiv:2208.10196 [hep-ph]].
%17 citations counted in INSPIRE as of 26 Mar 2024

\bibitem{Agaev:2022eyk}
S.~S.~Agaev, K.~Azizi and H.~Sundu,
%``Modeling the resonance Tcs0a(2900)++ as a hadronic molecule D*+K*+,''
Phys. Rev. D \textbf{107} (2023) no.9, 094019
doi:10.1103/PhysRevD.107.094019
[arXiv:2212.12001 [hep-ph]].
%10 citations counted in INSPIRE as of 26 Mar 2024

\bibitem{Ke:2022ocs}
H.~W.~Ke, Y.~F.~Shi, X.~H.~Liu and X.~Q.~Li,
%``Possible molecular states of D\textasciimacron{}*K* (D*K*) and new exotic states X0(2900), X1(2900), Tcs0a(2900)0 and Tcs0a(2900)++,''
Phys. Rev. D \textbf{106} (2022) no.11, 114032
doi:10.1103/PhysRevD.106.114032
[arXiv:2210.06215 [hep-ph]].
%17 citations counted in INSPIRE as of 01 Jan 2025

\bibitem{Duan:2023lcj}
M.~Y.~Duan, M.~L.~Du, Z.~H.~Guo, E.~Wang and D.~Y.~Chen,
%``Coupled-channel $D^\ast K^\ast -D_s^\ast \rho$ interactions and the origin of $T_{c\bar{s}0}(2900)$,''
Phys. Rev. D \textbf{108} (2023) no.7, 074006
doi:10.1103/PhysRevD.108.074006
[arXiv:2307.04092 [hep-ph]].
%14 citations counted in INSPIRE as of 01 Jan 2025

\bibitem{Wang:2023hpp}
B.~Wang, K.~Chen, L.~Meng and S.~L.~Zhu,
%``Spectrum of the molecular tetraquarks: Unraveling the Tcs0(2900) and Tcs\textasciimacron{}0a(2900),''
Phys. Rev. D \textbf{109} (2024) no.3, 034027
doi:10.1103/PhysRevD.109.034027
[arXiv:2309.02191 [hep-ph]].
%4 citations counted in INSPIRE as of 26 Mar 2024
%\cite{Faessler:2007gv}

\bibitem{Huang:2023fvj}
Y.~Huang, H.~Hei, J.~w.~Feng, X.~Chen and R.~Wang,
%``Production of the newly observed $\bar{T}_{c\bar{s}0}$ by kaon-induced reactions on a proton/neutron target,''
Phys. Rev. D \textbf{108} (2023) no.7, 076019
doi:10.1103/PhysRevD.108.076019
[arXiv:2308.14148 [hep-ph]].
%3 citations counted in INSPIRE as of 01 Jan 2025

\bibitem{Yu:2023avh}
Z.~Yu, Q.~Wu and D.~Y.~Chen,
%``$X_0(2900)$ and its spin partners productions in the $B^+$ decay processes,''
Eur. Phys. J. C \textbf{84} (2024) no.9, 985
doi:10.1140/epjc/s10052-024-13366-5
[arXiv:2310.12398 [hep-ph]].
%6 citations counted in INSPIRE as of 21 May 2025

\bibitem{Yang:2024coj}
Z.~Y.~Yang, Q.~Wang and W.~Chen,
%``Production and decay of the X0(2900) state with different interpretations,''
Phys. Rev. D \textbf{111} (2025) no.7, 076030
doi:10.1103/PhysRevD.111.076030
[arXiv:2412.02997 [hep-ph]].
%1 citations counted in INSPIRE as of 21 May 2025

\bibitem{Zhang:2020oze}
J.~R.~Zhang,
%``Open-charm tetraquark candidate: Note on $X_0$(2900),''
Phys. Rev. D \textbf{103} (2021) no.5, 054019
doi:10.1103/PhysRevD.103.054019
[arXiv:2008.07295 [hep-ph]].
%55 citations counted in INSPIRE as of 04 Jan 2025

\bibitem{Wang:2020xyc}
Z.~G.~Wang,
%``Analysis of the $X_0(2900)$ as the scalar tetraquark state via the QCD sum rules,''
Int. J. Mod. Phys. A \textbf{35} (2020) no.30, 2050187
doi:10.1142/S0217751X20501870
[arXiv:2008.07833 [hep-ph]].
%60 citations counted in INSPIRE as of 04 Jan 2025

\bibitem{Wang:2020prk}
G.~J.~Wang, L.~Meng, L.~Y.~Xiao, M.~Oka and S.~L.~Zhu,
%``Mass spectrum and strong decays of tetraquark ${\bar{c}}{\bar{s}} qq$ states,''
Eur. Phys. J. C \textbf{81} (2021) no.2, 188
doi:10.1140/epjc/s10052-021-08978-0
[arXiv:2010.09395 [hep-ph]].
%42 citations counted in INSPIRE as of 04 Jan 2025

\bibitem{He:2020jna}
X.~G.~He, W.~Wang and R.~Zhu,
%``Open-charm tetraquark $X_c$ and open-bottom tetraquark $X_b$,''
Eur. Phys. J. C \textbf{80} (2020) no.11, 1026
doi:10.1140/epjc/s10052-020-08597-1
[arXiv:2008.07145 [hep-ph]].
%69 citations counted in INSPIRE as of 04 Jan 2025

\bibitem{Guo:2021mja}
T.~Guo, J.~Li, J.~Zhao and L.~He,
%``Mass spectra and decays of open-heavy tetraquark states,''
Phys. Rev. D \textbf{105} (2022) no.5, 054018
doi:10.1103/PhysRevD.105.054018
[arXiv:2108.06222 [hep-ph]].
%19 citations counted in INSPIRE as of 04 Jan 2025

\bibitem{LHCb:2024iuo}
R.~Aaij \textit{et al.} [LHCb],
%``Study of $D_{s1}(2460)^{+}\to D_{s}^{+}\pi^{+}\pi^{-}$ in $B\to {\bar{D}}^{(*)}D_{s}^{+}\pi^{+}\pi^{-}$ decays,''
Sci. Bull. \textbf{70} (2025), 1432-1444
doi:10.1016/j.scib.2025.02.025
[arXiv:2411.03399 [hep-ex]].
%4 citations counted in INSPIRE as of 21 May 2025

\bibitem{Wang:2024fsz}
Z.~Y.~Wang, Y.~S.~Li and S.~Q.~Luo,
%``Scalar resonance contributions in the Ds1(2460)+\textrightarrow{}Ds+\ensuremath{\pi}+\ensuremath{\pi}- reaction,''
Phys. Rev. D \textbf{111} (2025) no.7, 076009
doi:10.1103/PhysRevD.111.076009
[arXiv:2412.06446 [hep-ph]].
%2 citations counted in INSPIRE as of 21 May 2025

\bibitem{Gregory:2025ium}
E.~B.~Gregory, F.~K.~Guo, C.~Hanhart, S.~Krieg and T.~Luu,
%``Exclusion of a diquark-antidiquark structure for the lightest positive-parity charmed mesons,''
[arXiv:2503.23954 [hep-lat]].
%0 citations counted in INSPIRE as of 14 May 2025

\bibitem{Kong:2021ohg}
S.~Y.~Kong, J.~T.~Zhu, D.~Song and J.~He,
%``Heavy-strange meson molecules and possible candidates Ds0*(2317), Ds1(2460), and X0(2900),''
Phys. Rev. D \textbf{104} (2021) no.9, 094012
doi:10.1103/PhysRevD.104.094012
[arXiv:2106.07272 [hep-ph]].
%27 citations counted in INSPIRE as of 26 Mar 2024

\bibitem{Yang:2023wgu}
G.~Yang, J.~Ping and J.~Segovia,
%``The tetraquark system in a chiral quark model*,''
Chin. Phys. C \textbf{48} (2024) no.7, 073106
doi:10.1088/1674-1137/ad39cd
[arXiv:2311.10376 [hep-ph]].
%2 citations counted in INSPIRE as of 21 May 2025



\bibitem{Chen:2023syh}
Y.~K.~Chen, W.~L.~Wu, L.~Meng and S.~L.~Zhu,
%``Unified description of the Qsq\textasciimacron{}q\textasciimacron{} molecular bound states, molecular resonances, and compact tetraquark states in the quark potential model,''
Phys. Rev. D \textbf{109} (2024) no.1, 014010
doi:10.1103/PhysRevD.109.014010
[arXiv:2310.14597 [hep-ph]].
%1 citations counted in INSPIRE as of 26 Mar 2024

\bibitem{Weinberg:1962hj}
S.~Weinberg,
  %``Elementary particle theory of composite particles,''
  Phys.\ Rev.\  {\bf 130}, 776 (1963).
  %%CITATION = PHRVA,130,776;%%
  %352 citations counted in INSPIRE as of 28 Sep 2015

\bibitem{Dong:2017gaw}
Y.~Dong, A.~Faessler and V.~E.~Lyubovitskij,
%``Description of heavy exotic resonances as molecular states using phenomenological Lagrangians,''
Prog. Part. Nucl. Phys. \textbf{94} (2017), 282-310
doi:10.1016/j.ppnp.2017.01.002
%85 citations counted in INSPIRE as of 26 Mar 2024

\bibitem{Guo:2017jvc}
F.~K.~Guo, C.~Hanhart, U.~G.~Mei{\ss}ner, Q.~Wang, Q.~Zhao and B.~S.~Zou,
%``Hadronic molecules,''
Rev. Mod. Phys. \textbf{90} (2018) no.1, 015004
[erratum: Rev. Mod. Phys. \textbf{94} (2022) no.2, 029901]
doi:10.1103/RevModPhys.90.015004
[arXiv:1705.00141 [hep-ph]].
%1437 citations counted in INSPIRE as of 29 Sep 2025
\bibitem{Xiao:2020ltm}
C.~J.~Xiao, D.~Y.~Chen, Y.~B.~Dong and G.~W.~Meng,
%``Study of the decays of $S-$wave $\bar D^\ast K^\ast$ hadronic molecules: The scalar $X_0(2900)$ and its spin partners $X_{J(J=1,2)}$,''
Phys. Rev. D \textbf{103} (2021) no.3, 034004
doi:10.1103/PhysRevD.103.034004
[arXiv:2009.14538 [hep-ph]].
%42 citations counted in INSPIRE as of 26 Mar 2024

\bibitem{Chen:2016byt}
D.~Y.~Chen, Y.~B.~Dong, M.~T.~Li and W.~L.~Wang,
%``Pionic transition from Y(4260) to Z$_{c}$(3900) in a hadronic molecular scenario,''
Eur. Phys. J. A \textbf{52} (2016) no.10, 310
doi:10.1140/epja/i2016-16310-0
%12 citations counted in INSPIRE as of 26 Mar 2024

\bibitem{Gutsche:2010jf}
T.~Gutsche, T.~Branz, A.~Faessler, I.~W.~Lee and V.~E.~Lyubovitskij,
%``Hadron Molecules,''
Chin. Phys. C \textbf{34} (2010) no.9, 1185-1190
doi:10.1088/1674-1137/34/9/007
[arXiv:1001.1870 [hep-ph]].
%7 citations counted in INSPIRE as of 01 Jan 2025

\bibitem{Salam:1962ap}
  A.~Salam,
  %``Lagrangian theory of composite particles,''
  Nuovo Cim.\  {\bf 25}, 224 (1962).
  %%CITATION = NUCIA,25,224;%%
  %183 citations counted in INSPIRE as of 28 Sep 2015
 %\cite{Hayashi:1967}
  
  \bibitem{Hayashi:1967}
  K. Hayashi, M.
    Hirayama, T. Muta, N. Seto, and T. Shirafuji, Fortschr.
    Phys. 15, 625 (1967).
%\cite{Kaymakcalan:1983qq}

\bibitem{vanKolck:2022lqz}
U.~van Kolck,
%``Weinberg's Compositeness,''
Symmetry \textbf{14} (2022), 1884
doi:10.3390/sym14091884
[arXiv:2209.08432 [hep-ph]].
%8 citations counted in INSPIRE as of 26 Mar 2024

\bibitem{Kaymakcalan:1983qq}
  O.~Kaymakcalan, S.~Rajeev and J.~Schechter,
  %``Nonabelian Anomaly and Vector Meson Decays,''
  Phys.\ Rev.\ D {\bf 30}, 594 (1984).
  %%CITATION = PHRVA,D30,594;%%
  %371 citations counted in INSPIRE as of 28 Sep 2015
 %\cite{Oh:2000qr}

\bibitem{Oh:2000qr}
  Y.~s.~Oh, T.~Song and S.~H.~Lee,
  %``J / psi absorption by pi and rho mesons in meson exchange model with anomalous parity interactions,''
  Phys.\ Rev.\ C {\bf 63}, 034901 (2001)
  [nucl-th/0010064].
  %%CITATION = NUCL-TH/0010064;%%
  %126 citations counted in INSPIRE as of 28 Sep 2015

\bibitem{Casalbuoni:1996pg}
  R.~Casalbuoni, A.~Deandrea, N.~Di Bartolomeo, R.~Gatto, F.~Feruglio and G.~Nardulli,
  %``Phenomenology of heavy meson chiral Lagrangians,''
  Phys.\ Rept.\  {\bf 281}, 145 (1997)
  [hep-ph/9605342].
  %%CITATION = HEP-PH/9605342;%%
  %397 citations counted in INSPIRE as of 28 Sep 2015
  %\cite{Colangelo:2002mj}

\bibitem{Colangelo:2002mj}
  P.~Colangelo, F.~De Fazio and T.~N.~Pham,
  %``B- ---> K- (chi(c0)) decay from charmed meson rescattering,''
  Phys.\ Lett.\ B {\bf 542}, 71 (2002)
  [hep-ph/0207061].
  %%CITATION = HEP-PH/0207061;%%
  %80 citations counted in INSPIRE as of 28 Sep 2015

\bibitem{He:2019csk}
J.~He, Y.~Liu, J.~T.~Zhu and D.~Y.~Chen,
%``Y(4626) as a molecular state from interaction ${D}^*_s{\bar{D}}_{s1}(2536)-{D}_s{\bar{D}}_{s1}(2536)$,''
Eur. Phys. J. C \textbf{80} (2020) no.3, 246
doi:10.1140/epjc/s10052-020-7820-2
[arXiv:1912.08420 [hep-ph]].
%22 citations counted in INSPIRE as of 01 Jan 2025

\bibitem{Ding:2008gr}
G.~J.~Ding,
%``Are Y(4260) and Z+(2) are D(1) D or D(0) D* Hadronic Molecules?,''
Phys. Rev. D \textbf{79} (2009), 014001
doi:10.1103/PhysRevD.79.014001
[arXiv:0809.4818 [hep-ph]].
%205 citations counted in INSPIRE as of 01 Jan 2025

\bibitem{Wu:2021udi}
Q.~Wu, D.~Y.~Chen and T.~Matsuki,
%``A phenomenological analysis on isospin-violating decay of $X(3872)$,''
Eur. Phys. J. C \textbf{81} (2021) no.2, 193
doi:10.1140/epjc/s10052-021-08984-2
[arXiv:2102.08637 [hep-ph]].
%30 citations counted in INSPIRE as of 01 Jan 2025

\bibitem{Liu:2020ruo}
J.~Liu, Q.~Wu, J.~He, D.~Y.~Chen and T.~Matsuki,
%``Production of $P-$wave charmed and charmed-strange mesons in pion and kaon induced reactions,''
Phys. Rev. D \textbf{101} (2020) no.1, 014003
doi:10.1103/PhysRevD.101.014003
[arXiv:2001.00212 [hep-ph]].
%8 citations counted in INSPIRE as of 01 Jan 2025

\bibitem{MARK-III:1988crp}
D.~Coffman \textit{et al.} [MARK-III],
%``Measurements of $J/\psi$ Decays Into a Vector and a Pseudoscalar Meson,''
Phys. Rev. D \textbf{38} (1988), 2695
[erratum: Phys. Rev. D \textbf{40} (1989), 3788]
doi:10.1103/PhysRevD.38.2695
%128 citations counted in INSPIRE as of 01 Jan 2025

\bibitem{Morgan:1970yz}
D.~L.~Morgan and V.~W.~Hughes,
%``Atomic processes involved in matter-antimatter annihilation,''
Phys. Rev. D \textbf{2} (1970), 1389-1399
doi:10.1103/PhysRevD.2.1389
%37 citations counted in INSPIRE as of 01 Jan 2025

\bibitem{Liu:2005jb}
  W.~Liu, C.~M.~Ko and L.~W.~Chen,
  %``Eta absorption by mesons,''
  Nucl.\ Phys.\ A {\bf 765} (2006) 401
  [nucl-th/0505075].
  %%CITATION = NUCL-TH/0505075;%%
  %4 citations counted in INSPIRE as of 09 sept. 2015
%\cite{Bando:1985rf}

\bibitem{Bando:1985rf}
  M.~Bando, T.~Kugo and K.~Yamawaki,
  %``On the Vector Mesons as Dynamical Gauge Bosons of Hidden Local Symmetries,''
  Nucl.\ Phys.\ B {\bf 259}, 493 (1985).
  %%CITATION = NUPHA,B259,493;%%
  %298 citations counted in INSPIRE as of 28 Sep 2015

\bibitem{Chen:2011cj}
D.~Y.~Chen, X.~Liu and T.~Matsuki,
%``Two Charged Strangeonium-Like Structures Observable in the $Y(2175) \to \phi(1020)\pi^{+} \pi^{-}$ Process,''
Eur. Phys. J. C \textbf{72} (2012), 2008
doi:10.1140/epjc/s10052-012-2008-z
[arXiv:1112.3773 [hep-ph]].
%38 citations counted in INSPIRE as of 01 Jan 2025

\bibitem{Haglin:2000ar}
K.~L.~Haglin and C.~Gale,
%``Hadronic interactions of the J / psi,''
Phys. Rev. C \textbf{63} (2001), 065201
doi:10.1103/PhysRevC.63.065201
[arXiv:nucl-th/0010017 [nucl-th]].
%119 citations counted in INSPIRE as of 01 Jan 2025

\bibitem{Lin:1999ad}
Z.~w.~Lin and C.~M.~Ko,
%``A Model for J / psi absorption in hadronic matter,''
Phys. Rev. C \textbf{62} (2000), 034903
doi:10.1103/PhysRevC.62.034903
[arXiv:nucl-th/9912046 [nucl-th]].
%255 citations counted in INSPIRE as of 01 Jan 2025

\bibitem{Baru:2003qq}
V.~Baru, J.~Haidenbauer, C.~Hanhart, Y.~Kalashnikova and A.~E.~Kudryavtsev,
%``Evidence that the a(0)(980) and f(0)(980) are not elementary particles,''
Phys. Lett. B \textbf{586}, 53-61 (2004)
doi:10.1016/j.physletb.2004.01.088
[arXiv:hep-ph/0308129 [hep-ph]].
%475 citations counted in INSPIRE as of 28 Sep 2025

%\cite{Lin:2017mtz}
\bibitem{Lin:2017mtz}
Y.~H.~Lin, C.~W.~Shen, F.~K.~Guo and B.~S.~Zou,
%``Decay behaviors of the $P_c$ hadronic molecules,''
Phys. Rev. D \textbf{95} (2017) no.11, 114017
doi:10.1103/PhysRevD.95.114017
[arXiv:1703.01045 [hep-ph]].
%69 citations counted in INSPIRE as of 28 Sep 2025


\bibitem{Aubert:2006mh}
  B.~Aubert {\it et al.} [BaBar Collaboration],
  %``Observation of a New D(s) Meson Decaying to DK at a Mass of 2.86-GeV/c**2,''
  Phys.\ Rev.\ Lett.\  {\bf 97}, 222001 (2006)
  [hep-ex/0607082].
  %%CITATION = HEP-EX/0607082;%%
  %161 citations counted in INSPIRE as of 25 Sep 2015
 %\cite{Agashe:2014kda}

\bibitem{Belyaev:1994zk}
V.~M.~Belyaev, V.~M.~Braun, A.~Khodjamirian and R.~Ruckl,
%``D* D pi and B* B pi couplings in QCD,''
Phys. Rev. D \textbf{51} (1995), 6177-6195
doi:10.1103/PhysRevD.51.6177
[arXiv:hep-ph/9410280 [hep-ph]].
%524 citations counted in INSPIRE as of 01 Jan 2025

%\cite{Baru:2003qq}
\bibitem{LHCb:2024vfz}
R.~Aaij \textit{et al.} [LHCb],
%``Observation of New Charmonium or Charmoniumlike States in B+\textrightarrow{}D*\ensuremath{\pm}D\ensuremath{\mp}K+ Decays,''
Phys. Rev. Lett. \textbf{133} (2024) no.13, 131902
doi:10.1103/PhysRevLett.133.131902
[arXiv:2406.03156 [hep-ex]].
%14 citations counted in INSPIRE as of 31 Dec 2024
%\cite{ParticleDataGroup:2024cfk}
\end{thebibliography}
\end{document}